\documentclass[
 reprint,
superscriptaddress,twocolumn,
 amsmath,amssymb,
 aps,
]{revtex4-2}

\usepackage{mathtools}
\usepackage{graphicx}
\usepackage{dcolumn}
\usepackage{bm}
\usepackage{tensor}
\usepackage{physics}
\usepackage{url}

\begin{document}

\preprint{APS/123-QED}

\title{Dual epitaxial telecom spin-photon interfaces with correlated long-lived coherence}

\author{Shobhit Gupta}
\affiliation{Department of Physics, University of Chicago, Chicago IL 60637 USA}

\author{Yizhong Huang}
\affiliation{Pritzker School of Molecular Engineering, University of Chicago, Chicago IL 60637 USA}

\author{Shihan Liu}
\affiliation{Pritzker School of Molecular Engineering, University of Chicago, Chicago IL 60637 USA}

\author{Yuxiang Pei}
\affiliation{Department of Physics, University of Chicago, Chicago IL 60637 USA}

\author{Natasha Tomm}
\affiliation{Department of Physics, University of Basel, 4056 Basel, Switzerland}

\author{Richard J. Warburton}
\affiliation{Department of Physics, University of Basel, 4056 Basel, Switzerland}

\author{Tian Zhong}
\email{tzh@uchicago.edu}
\affiliation{Pritzker School of Molecular Engineering, University of Chicago, Chicago IL 60637 USA}

\begin{abstract}
Optically active solid-state spin qubits thrive as an appealing technology for quantum interconnect and quantum networking, owing to their atomic size, scalable creation, long-lived coherence, and ability to coherently interface with flying qubits. Trivalent erbium dopants, in particular, emerge as a compelling candidate with their telecom C band emission and shielded 4f intra-shell spin-optical transitions. However, prevailing top-down architecture for rare-earth qubits and devices has yet allowed simultaneous long optical and spin coherence necessary for long-distance quantum networks. Here we demonstrate dual erbium telecom spin-photon interfaces in an epitaxial thin-film platform via wafer-scale bottom-up synthesis. Harnessing precise controls over the matrix purity, dopant placement, and symmetry unique to this platform, we simultaneously achieve millisecond erbium spin coherence times and $<$3 kilohertz optical dephasing rate in an inversion-symmetry protected site and realize both optical and microwave control in a fiber-integrated package for rapid scaling up. These results demonstrate a significant prospect for high-quality rare-earth qubits and quantum memories assembled using a bottom-up method and pave the way for the large-scale development of quantum light-matter interfaces for telecommunication quantum networks. 
\end{abstract}

\maketitle

\raggedbottom

\noindent 
Generation of entanglement over long-distance optical network \cite{Kimble2008, PRXQuantum.2.017002, doi:10.1126/science.aam9288} underscores multitudes of quantum information applications in secure communication, distributed quantum sensing, and quantum computation, and requires quantum light-matter interfaces as a dispensable building block. Such interfaces can be realized with spin qubits in individual atoms with a photon-emitting optical transition, preferably in low-loss telecommunication bands. Solid-state spin-photon interfaces such as quantum dots \cite{Delteil2016, PhysRevLett.119.010503, PhysRevLett.110.167401,DeGreve2012, Gao2012,PhysRevLett.128.233602}, defects in diamond \cite{Bar-Gill2013, Bernien2013, Hensen2015, Hermans2022, doi:10.1126/science.add9771, Bhaskar2020} and silicon carbide \cite{chrisdoi:10.1126/sciadv.abm5912}, T centers \cite{Higginbottom2022, PRXQuantum.1.020301, PRXQuantum.4.020308}, and rare-earth ions are among the most promising candidates to date. However, few platforms so far have simultaneously demonstrated all desired properties, namely, long qubit coherence times exceeding a millisecond (for the practical network over 100km \cite{PhysRevA.80.032301}), coherent photon emission with ideally transform-limited linewidths in the telecom C- or O-bands, and scalable device integration allowing efficient channeling of emission into optical fibers. For instance, trivalent erbium dopants in crystals possess a telecom C band emission and have been investigated extensively for quantum memories and repeaters \cite{Asadi_2020, KimiaeeAsadi2018quantumrepeaters}. Milliseconds Er spin coherence times \cite{doi:10.1126/sciadv.abj9786, y2o3millisecond2022}, narrow optical linewidths \cite{erbiumSDpaper, yizhongandriku, PhysRevX.12.041009}, coherent control and readout of single Er spins \cite{Raha2020, ourari2023indistinguishable}, have all been demonstrated but separately in different host crystals and device configurations. An ensuing challenge is to develop a unified Er qubit platform that enables simultaneous long optical and spin coherence with plug-play deployability in fiber-optic telecommunication networks. 

Engineering designer properties of rare-earth spin-photon interface boils down to atomic design, synthesis, and control of the local matrix environment surrounding individual dopants using fundamental principles of symmetry, ligand field theory, and crystal kinetics. The prevailing material platform based on bulk single crystals \cite{THIEL2011353} exhibit supreme crystalline qualities as evidenced by record coherence lifetimes in $\mathrm{Y_2SiO_5}$ \cite{erbiumSDpaper, Ortu2018, Rancic2018}, $\mathrm{Y_2O_3}$ \cite{y2o3millisecond2022, y2o3esr2022, yizhongandriku, Zhang2017}, $\mathrm{YVO_4}$ \cite{Kindem2020, Ruskuc2022}, $\mathrm{CaWO_4}$ \cite{wang2023single, doi:10.1126/sciadv.abj9786}, LiNbO$_3$ \cite{PhysRevApplied.18.014069, Xia:22, Yang2023} at cryogenic temperatures. However, current growth techniques of these crystals render minimal nanoscopic control. Top-down synthesis like ion implantation \cite{PhysRevApplied.19.014037, PhysRevX.12.041009, Weiss:21, Gritsch:23,10.1063/5.0046904, PhysRevApplied.18.014069} and photonic integration using bulk crystals have led to successful optical addressing of single ions \cite{PRLsinglesTian, PhysRevLett.120.243601, yu2023frequency, Xia:22, 10.108870301, Yang2023, horvath2023strong, doi:10.1126/sciadv.abo4538, Gritsch:23, Xia:22, Yang2023}, spectral multiplexing \cite{doi:10.1126/sciadv.abo4538} and nuclear spin control\cite{Ruskuc2022, PRXQuantum.4.010323}. However, varying degrees of undermined coherence for dopants embedded in these devices compared with their bulk crystal counterpart have raised questions about the possible impact of device fabrication and proximity to surfaces \cite{wu2022hybrid, PhysRevX.10.041025}. 

To overcome limitations in conventional top-down approaches, here we develop a wafer-scale rare-earth qubit platform using bottom-up epitaxial growth of Er dopants in $\mathrm{Y_2O_3}$ thin film single crystals, showcasing an unprecedented control of matrix purity, precision dopant placement, device integrability, and scalability \cite{nanoREreview}. Exploiting two distinct symmetry sites of the $\mathrm{Y_2O_3}$ lattice, we perform correlated spin-optical spectroscopy by coupling Er qubits to both superconducting microwave and fiber micro-resonators. Site-resolved spin addressing and symmetry protection enable us to engineer a simultaneous millisecond-long spin coherence and kilohertz optical coherence for erbium qubits in an inversion-symmetric lattice site. These qubits are packaged in a fully fiber-compatible device architecture, and are readily scaled up for mass production, thus representing a viable path towards deployable quantum interconnect technologies with uncompromised coherence performance for telecommunication quantum networks.\newline

\begin{figure*}[th]
    \centering
    \includegraphics[width=1\textwidth]{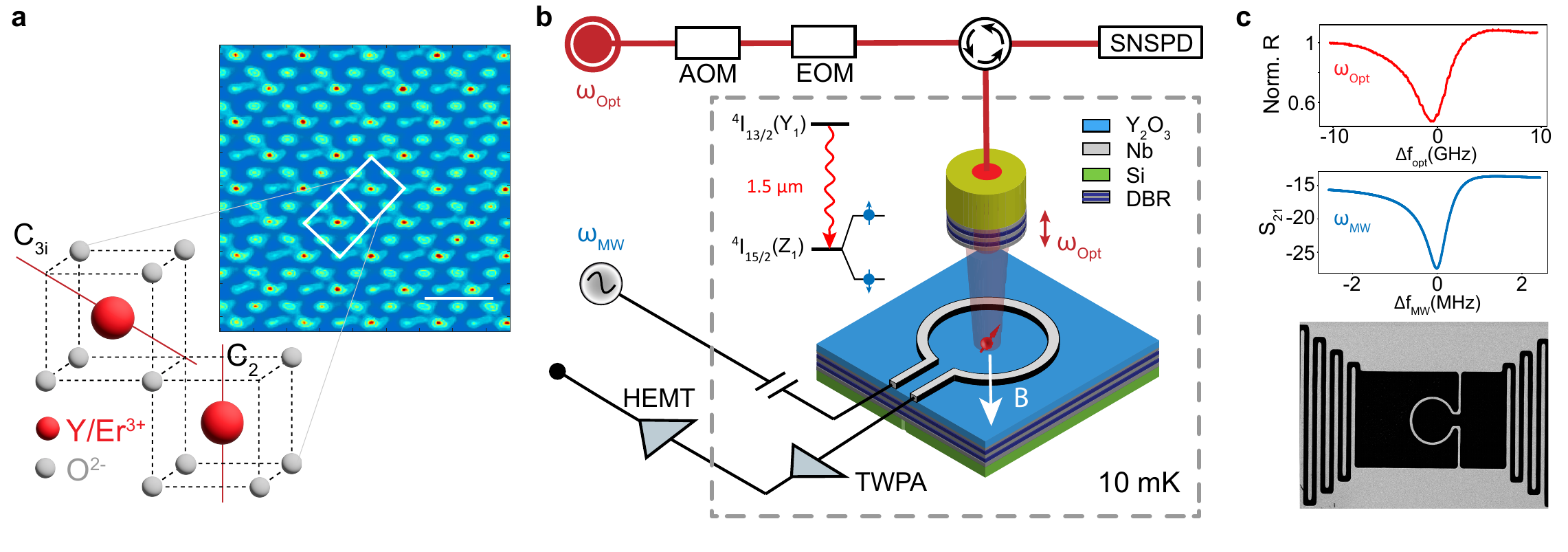}
    \caption{ {\bf Schematics of the Er$^{3+}$ qubit device and experiment setup.} {\bf a}. Transmission electron microscope (TEM) lattice image of the cubic-phase epitaxial $\mathrm{Y_2 O_3}$ films grown on silicon. The scale bar is 1 nm. Yttrium oxide lattice possesses a low symmetry C\textsubscript{2} and a high symmetry C\textsubscript{3i} site \cite{Som2013}, with their respective symmetry axes shown in red. {\bf b}. The device consists of a tunable cryogenic fiber Fabry-Perot cavity coupled to $\mathrm{^{4}I_{15/2} \, Z_1} \rightarrow \mathrm{^{4}I_{13/2}} \, \mathrm{Y_1}$ optical transition of $\mathrm{Er^{3+}}$ and a low-impedance superconducting microwave resonator coupled to $\mathrm{Er^{3+}}$ spin transition at 5-6 GHz. {\bf c}. (Top) the reflection spectrum of the fiber Fabry-Perot cavity shows a Q of $\mathrm{58,000}$. (Middle) spectrum of the microwave resonator exhibiting an intrinsic Q of $\mathrm{370,000}$ and an estimated single spin coupling strength $\mathrm{\gtrapprox \,100 \, Hz}$. (Bottom) SEM image of a co-planar superconducting microwave resonator based on niobium. SNSPD: superconducting nanowire single photon detector; TWPA: traveling wave parametric amplifier; HEMT: high electron mobility transistor. DBR: distributed Bragg reflector}
    \label{fig1}
\end{figure*}

{\noindent \bf Fiber-integrated Er qubit chip}
\noindent Figure~\ref{fig1} illustrates the schematics of our Er$^{3+}$ qubit device. A die chip consists of a sub-wavelength thick, cubic-phase $\mathrm{Er^{3+}}$ doped $\mathrm{Y_2O_3}$ film that is transferred onto a distributed Bragg reflector (DBR) stack on a silicon substrate (See Methods). The high-purity, single crystal Er$^{3+}$:Y$_2$O$_3$ films were epitaxially grown using molecular beam epitaxy (MBE) on silicon (111) wafers \cite{doi:10.1063/1.5142611}. The Er dopants are placed at least 40 nm from the top and bottom crystal interfaces during the layer-by-layer growth. $\mathrm{Er^{3+}}$ substitutes $\mathrm{Y^{3+}}$ in two lattice sites: a $\mathrm{C_2}$ site with both electric and magnetic dipole allowed $\mathrm{Z_1} \rightarrow \mathrm{Y_1}$ optical transition, and a $\mathrm{C_{3i}}$ site with only magnetic dipole allowed optical transition due to quenched permanent electric dipole by the inversion symmetry (Fig.~\ref{fig1}a).

Vertical to the plane of the chip, a tunable fiber Fabry-Perot cavity formed by the DBR substrate and a DBR-coated dimpled fiber\cite{10.1063/1.3679721} couples to the 1.5 $\mathrm{\mu m}$ telecom-C band transitions between the lowest crystal field doublets $\mathrm{Z_1}$ and $\mathrm{Y_1}$ of the $\mathrm{^4I_{15/2}}$ and $\mathrm{^4I_{13/2}}$ spin-orbit levels of Er$^{3+}$ respectively. When the fiber tip is in rigid contact with the chip \cite{Najer2019}, the optical cavity reaches a smallest cavity length of (3/2)$\lambda$ (See Methods) with measured quality factors in a range of 45,000-60,000. A niobium co-planar superconducting microwave resonator with a low-impedance design to concentrate magnetic field inside an inductor loop was patterned on the Y$_2$O$_3$ layer to allow coupling to the microwave electron spin transitions between the Zeeman doublets of the $\mathrm{Z_1}$ optical ground state. The microwave resonator frequency is around 5-6 GHz with a typical measured Q of 1,500-3,000, and a single Er$^{3+}$ spin coupling strength $\mathrm{\geq}$100 Hz. The double optical-microwave resonators enable correlated spin-optical spectroscopy and control of the Er$^{3+}$ qubits. Combined with dual symmetry sites of the Er$^{3+}$, this platform makes possible a comprehensive mapping of electromagnetic noise in the epitaxial film matrix.\newline

\begin{figure*}[t!]
\centering 
\includegraphics[width=0.85\textwidth]{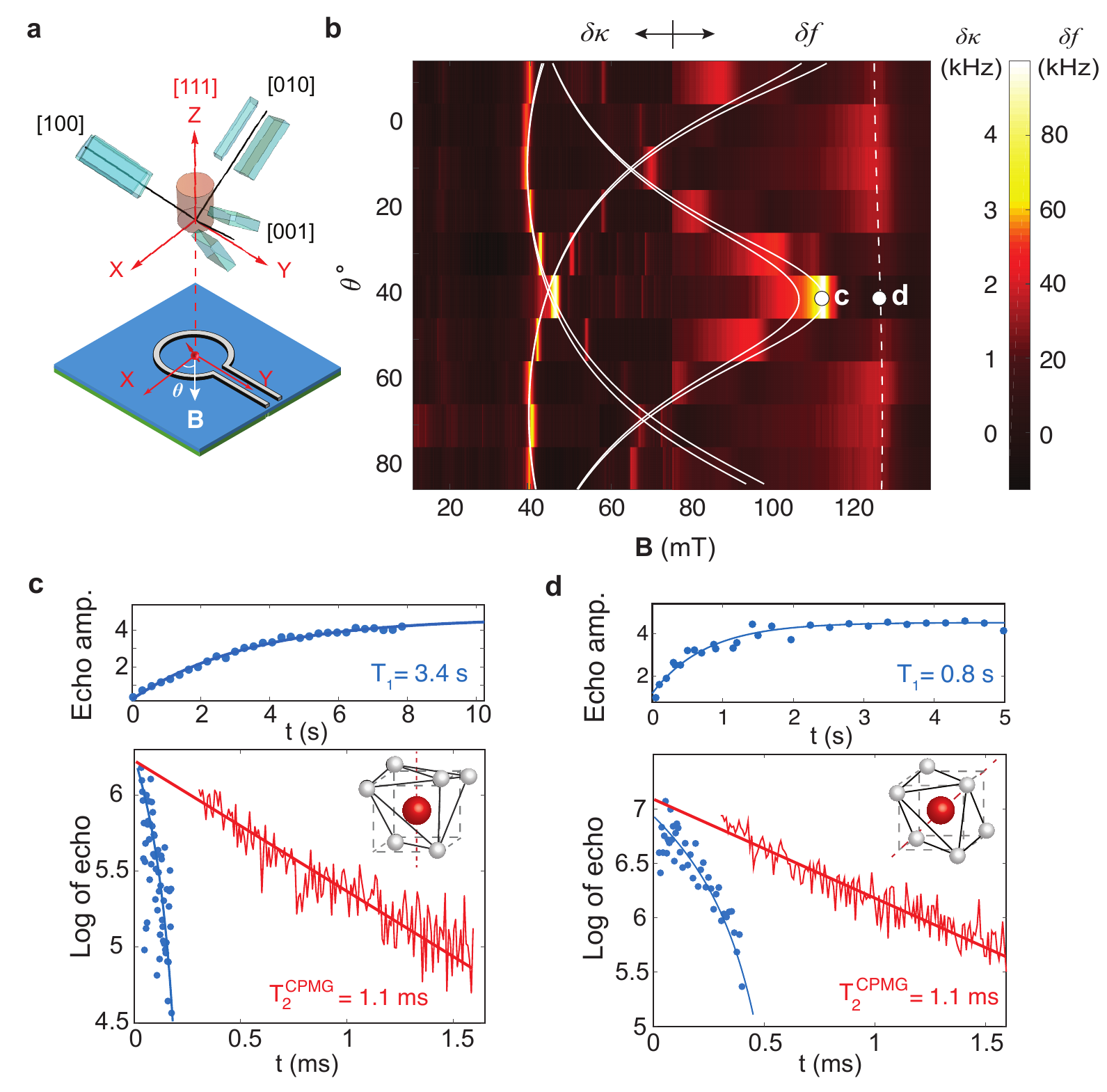}
\caption{ {\bf $\mathrm{Er}^{3+}$ spin anisotropy and coherence lifetimes in single-crystal $\mathrm{Y_2 O_3}$ thin films.} {\bf a}. Magnetic g-tensor axial orientation and anisotropy of six sub-sites of the $\mathrm{C_2}$ symmetry group are represented by cuboids, and one sub-site of the $\mathrm{C_{3i}}$ group as a cylinder, with Z axis along the crystallographic [111] direction of $\mathrm{Y_2 O_3}$. {\bf b}. Scanning the B field in the plane of the chip (XY plane) reveals anisotropic (dispersive and absorptive) coupling of $\mathrm{Er}^{3+}$ in distinct sub-sites to the microwave resonator. A close agreement with the theoretical model (white solid and dashed curves) for all $\mathrm{C_2}$ and $\mathrm{C_{3i}}$ sub-sites verifies the single crystallinity of the film. {\bf c}. Spin lifetime ($\mathrm{T_1}$ =3.4 s) (top) and coherence times (bottom) measured with Hahn echo ($\mathrm{T_{2}^{Hahn}}$=0.18 ms, blue) and Carr-Purcell-Meiboom-Gill (CPMG) sequence ($\mathrm{T_{2}^{CPMG}}$= 1.11 ms, red) for the $\mathrm{C_2}$ site g = 3.6 transition.  {\bf d}. Spin lifetime ($\mathrm{T_1}$=0.8 s) (top) and coherence times (bottom) measured with Hahn echo ($\mathrm{T_{2}^{Hahn}}$=0.38 ms, blue) and CPMG sequence ($\mathrm{T_{2}^{CPMG}}$= 1.14 ms, red) for the $\mathrm{C_{3i}}$ site g = 3.2 transition.} 
\label{fig2}
\end{figure*}

\noindent {\bf Er$^{3+}$ spin anisotropy and coherence lifetimes}
\noindent To probe the magnetic anisotropy of the Er$^{3+}$ spin qubits, we sweep the magnetic field intensity and the in-plane (XY) field angle $\mathrm{\theta}$ while detecting the electron spin resonance (ESR). Substitutional $\mathrm{Er^{3+}}$ ions in the $\mathrm{C_2}$ site occupy six orientationally inequivalent sub-sites with pairs of sub-sites sharing a symmetry axis (principal axis of the g-tensor) along the crystallographic [1,0,0], [0,1,0] and [0,0,1] directions (cuboids in Fig.~\ref{fig2}a). One sub-site of the $\mathrm{C_{3i}}$ site has rotational symmetry in the plane of the film (cylinder in Fig.~\ref{fig2}a). Coupling of $\mathrm{Er^{3+}}$ spins in each sub-site results in an absorptive broadening of the resonator linewidth $\mathrm{\delta \kappa}$ and a dispersive frequency shift $\mathrm{\delta f}$ by \cite{resonatorysoPhysRevApplied.11.054082}

\begin{equation}{\delta \kappa = \Omega^2 \gamma_s/ (\gamma_s^2+ \Delta^2)}\end{equation}
\noindent and
\begin{equation}{\delta f = - \Omega^2 \Delta/ (\gamma_s^2+ \Delta^2)},
\end{equation}

\noindent
where $\mathrm{\delta \kappa}$ is the increase in resonator linewidth, $\mathrm{\delta}f$ is the dispersive resonator frequency shift, $\mathrm{\Omega}$ is the spin ensemble coupling strength, $\mathrm{\gamma_s}$ is the spin inhomogeneous half-width, and $\mathrm{\Delta = g \mu_B  (B- B_0)/\hbar}$ with B$_0$ the resonant field. 

Figure ~\ref{fig2}b reveals the absorptive (dispersive) ESR signal for low (high) field transitions at different in-plane angles $\mathrm{\theta}$ (SI Fig.~S4). We clearly resolve three sets of $\mathrm{C_2}$ site spin transitions with a strong angular dependence and a $\mathrm{C_{3i}}$ site transition with nearly no angular dependence. These features are in good agreement with simulated resonance fields (white solid and dashed curves) calculated from the g-tensors of $\mathrm{Er^{3+}}$ in $\mathrm{Y_2 O_3}$ \cite{y2o3esr2022}, confirming the coupling of $\mathrm{Er^{3+}}$ spins. More importantly, the signature anisotropy of all symmetry sites proves of the single crystallinity of the $\mathrm{Y_2 O_3}$ film over the device area of 50 $\times$ 50 $\mu$m$^2$. Fitting the ESR signal gives a narrow spin inhomogeneous linewidth of 68 MHz, which is also found to be field angle-dependent (SI Section 1.3). Furthermore, we infer from the dispersive ESR signal a density of $\mathrm{Er^{3+}}$ spins of 13.6 parts per million (ppm) for each $\mathrm{C_{2}}$ sub-site and 12.4 ppm for the $\mathrm{C_{3i}}$ sub-site (SI Section 1.2).

Resolved Er sub-sites with distinct anisotropy enable us to enhance the qubit coherence times by operating at optimal field configurations. Specifically, spin transitions with lower g-factors, thus less noise sensitivity and strongly suppressed noise by a higher field, are preferred \cite{y2o3millisecond2022}. Fixing the magnetic field at $\mathrm{\theta}$= 40$^{\circ}$ (Fig.~\ref{fig2}b), we choose a $\mathrm{C_2}$ sub-site with g = 3.6 spin transition at 113 mT (circle {\bf c}), and a $\mathrm{C_{3i}}$ sub-site with g = 3.2 transition at 130 mT (circle {\bf d}) as target spin qubits. At 8.5 mK temperature, the $\mathrm{C_2}$ spins measured a two-pulse (Hahn) echo spin coherence time $\mathrm{T_{2}^{Hahn}}$ of 0.18 ms following a stretched exponential decay $\mathrm{e^{-(2t/(T_2))^n}}$ with a stretch factor n = 1.18, which indicates spectral diffusion as a source of decoherence \cite{mimsspectral}. We applied Carr-Purcell-Meiboom-Gill (CPMG) dynamical decoupling sequence \cite{PhysRev.94.630carlpurcell, doi:10.1063/1.1716296cpmg, dynamicdecouplingcolloq} with $N$ = 500 $\pi$ pulses and a pulse separation $\mathrm{2t=}$ 8 $\mathrm{\mu s}$ to suppress the spectral diffusion and obtain a $\mathrm{T_{2}^{CPMG}}$ of 1.11 ms, which we believe is limited by Er-Er spin interaction \cite{HARBRIDGE200344}(SI Section 1.7). Using XY8 sequence obtained comparable result of $\mathrm{T_{2}^{XY8}}=$ 0.9 ms (SI Section 1.6). The $\mathrm{C_{3i}}$ spins measured a longer Hahn echo spin coherence time $\mathrm{T_{2}^{Hahn}}$ of 0.38 ms with a stretch factor n = 2.11 and $\mathrm{T_{2}^{CPMG}}$ of 1.14 ms. Long spin relaxation times $\mathrm{T_1}$ = 3.4 s for $\mathrm{C_2}$ and 0.8 s for $\mathrm{C_{3i}}$ spins was measured using saturation recovery sequences. $\mathrm{T_1}$ in both sites show weak dependence on temperature and spin-cavity frequency detuning, indicating spin relaxations are dominated by direct-phonon process with a small contribution from spin flip-flops \cite{modifcationofphonon} (SI Section 1.8, 1.9). Considering the B$^5$ scaling of the direct phonon limited decay, we note that the spin $\mathrm{T_1}$ measured here are consistent with the longer (45 s) Er lifetimes reported at a much lower field and spin transition frequency \cite{Raha2020}\newline.

\noindent {\bf Wafer-scale coherence and noise spectroscopy}

\begin{figure*}[t!]
    \centering
    \includegraphics[width=0.85\textwidth]{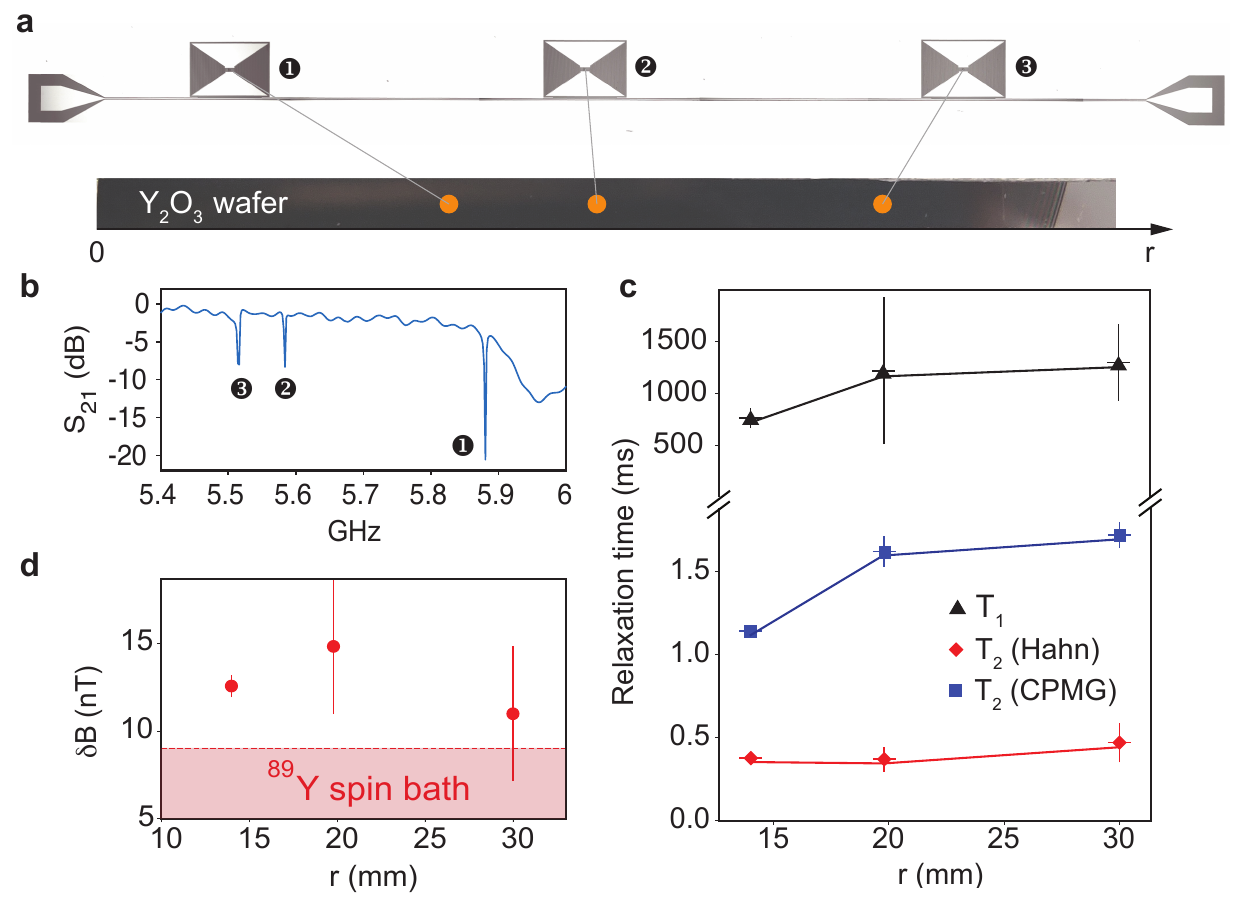}
    \caption{{\bf Spin coherence and noise spectroscopy on a wafer scale.} {\bf a}. Die chips were taken from different radial positions (r) with respect to the center of a $\mathrm{Er^{3+}}:\mathrm{Y_2 O_3}$ wafer, and flip-mounted individually on spectrally offset resonators coupled to a common transmission line. {\bf b}. The transmission spectrum of the devices shows three distinct device resonances. {\bf c}. Hahn echo (red) and CPMG (blue ) spin coherence times and spin lifetimes (black triangles) measurements. {\bf d}. Fluctuating magnetic noise amplitude ($\mathrm{\delta B}$) plotted as a function of distance from the center. Low magnetic noise amplitude $\mathrm{\delta B}$=11-13 nT, close to the yttrium nuclear spin limit, is measured on a wafer scale. Increasing $\mathrm{T_2^{CPMG}}$ and $\mathrm{T_1}$ with r indicates a reduced spin-spin interaction near the edges of the wafer due to decreased spin spectral density.}
    \label{fig:waferscale}
\end{figure*}

\noindent
The long-lived coherence of the Er qubit chip can be scaled up to the full $\mathrm{Y_2 O_3}$ wafer. We performed pulsed ESR on sample chips diced from different radial positions (r) from a $\mathrm{Er^{3+}}:\mathrm{Y_2 O_3}$ wafer (Fig.~\ref{fig:waferscale}a). The chips were flip-mounted on three superconducting resonators coupled to a common transmission line (Methods), and the resonators were offset in the frequency as shown in the measured transmission spectrum (Fig.~\ref{fig:waferscale}b). Spin coherence was measured on the $\mathrm{C_{3i}}$ site g =3.2 transition across different chips using the identical microwave pulses. We measured long spin $\mathrm{T_{2}^{Hahn}}$ coherence times in the range of 0.38-0.47 ms, indicating a consistently low magnetic noise, hence high purity of the host matrix over the wafer scale. Suppressing spectral diffusion using $N$ = 500 CPMG pulse sequence with a separation $\mathrm{2\tau}$= 8 $\mathrm{\mu s}$ achieved $\mathrm{T_2^{CPMG}}$ coherence times in the range of 1.14 - 1.72 ms (blue data in Fig.~\ref{fig:waferscale}d), which showed a modest improvement with increasing r (Extended data, SI Fig.~S8). From the measured $\mathrm{T_{2}^{Hahn}}$, $\mathrm{T_2^{CPMG}}$ we deduce a root-mean-square (RMS) fluctuating magnetic noise amplitude $\mathrm{\delta B}$ between 11-13 nT at different locations on the wafer (Fig.~\ref{fig:waferscale}c and SI Section 1.4), which is close to the yttrium nuclear spin-bath limit of $\mathrm{\leq}$ 9 nT in $\mathrm{Y_2 O_3}$ \cite{y2o3millisecond2022} and further evidences the high quality of the epitaxial film. The measured spin lifetimes $\mathrm{T_1}$ ranged from 0.76  to 1.30 s with increasing r, which we attribute to a decreased cross-relaxation between resonant spins for chips farther away from the center of the wafer due to a reduced spin spectral density (SI Section 1.7, Fig.~S9). This trend is supported by a measured broadening of spin inhomogeneous linewidth for chips near the edge of the wafer (SI Fig.~S10) and suggests that both $\mathrm{T_1}$ and $\mathrm{T_2^{CPMG}}$ can be improved by lowering $\mathrm{Er^{3+}}$ doping density, for instance, down to parts per billion level \cite{doi:10.1126/sciadv.abj9786}.\newline

\noindent {\bf Dual Er$^{3+}$ telecom spin-photon interface and optical coherence}
\begin{figure*}[t!]
    \centering
    \includegraphics[width=1.0\textwidth]{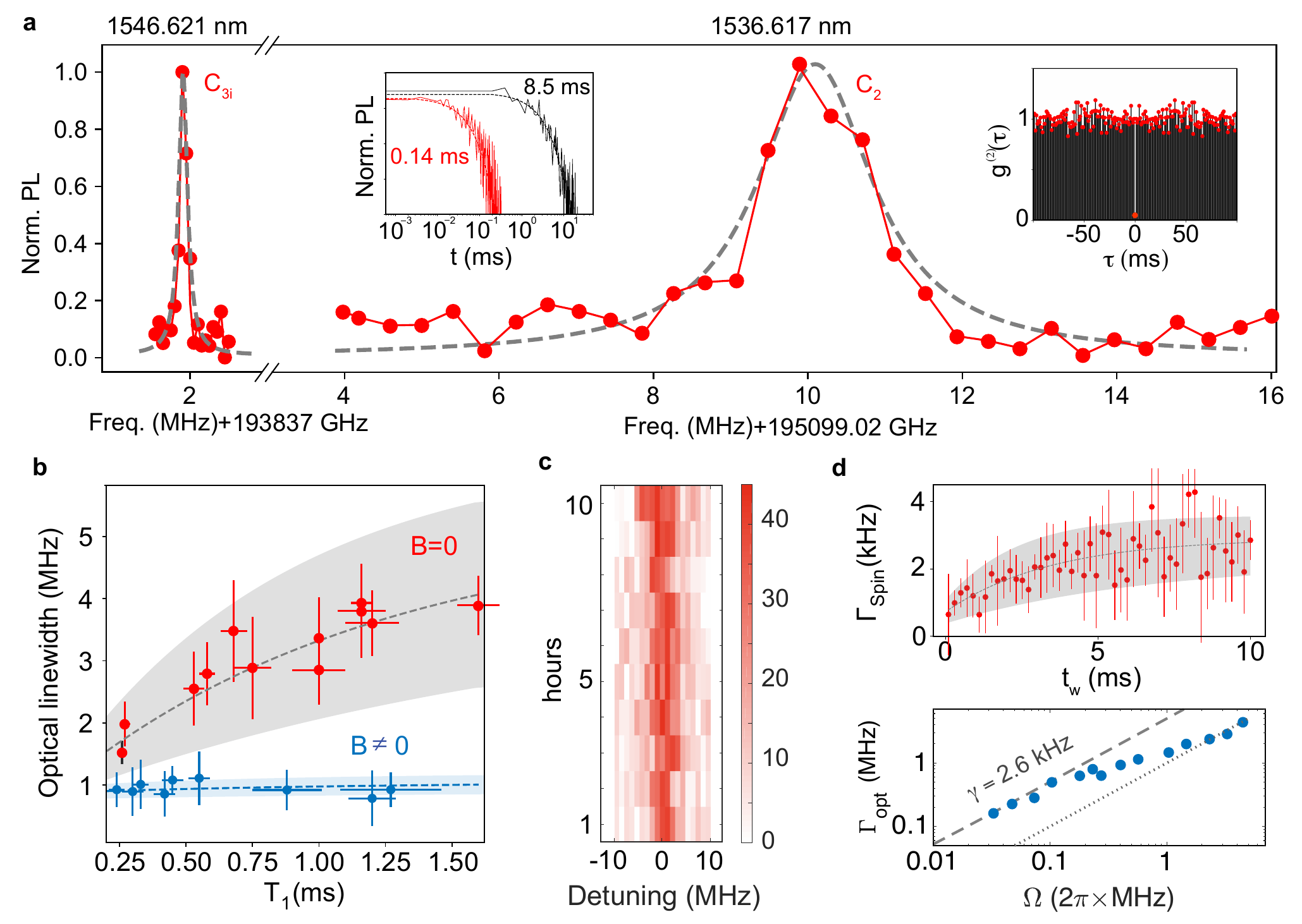}
    \caption{{\bf Dual Er spin-photon interface and optical coherence.} {\bf a}. Narrow photoluminescence excitation spectrum is observed from single $\mathrm{Er}^{3+}$ ions in $\mathrm{C_2}$ and $\mathrm{C_{3i}}$ site of $\mathrm{Y_2 O_3}$. (Right inset) $\mathrm{g^2}$ auto-correlation measurement for an $\mathrm{Er}^{3+}$ ion in C$_2$ site shows $\mathrm{g^2 (0)}= 0.06 $. (Left inset) Purcell enhancement of the same ion coupled to the optical cavity. {\bf b}. Short-term optical spectral diffusion of $\mathrm{Er}^{3+}$ in $\mathrm{C_{2}}$ site as optical $\mathrm{T_1}$ lifetimes are tuned with (blue, B = 250 mT) and without (red) applied B field. Grey(blue) dashes and shades indicate the fit and one standard deviation. {\bf c}. Long-term spectral stability of an $\mathrm{Er}^{3+}$ emitter in $\mathrm{C_2}$ site. {\bf d}. Correlated spin-optical coherence in $\mathrm{C_{3i}}$ site. (Top) Spin spectral diffusion linewidth with the same in-plane B (=130 mT) field as in Fig.~2(d), showing a maximum spin linewidth of 2.9 kHz. (Bottom) Optical homogeneous linewidth for the same site $\mathrm{Er}^{3+}$ under the same B field as a function of the optical excitation Rabi frequency. A minimum optical Rabi linewidth of 161 kHz was measured, corresponding to a pure dephasing rate of 2.6 kHz and a $T_2^*$ = 122 $\pm$ 5 $\mu$s (grey dashed line).}
    \label{fig4}
\end{figure*}
\noindent
A quantum spin-photon interface requires a spin-selective, coherent optical transition. Here we demonstrate the optical addressing of individual Er qubits coupled to a cryogenic fiber Fabry-Perot cavity (Methods) under the identical spin configurations in the previous experiments. Figure~\ref{fig4}a plots typical photoluminescence excitation (PLE) spectra of single Er ions in dual symmetry sites as the cavity resonance frequency is scanned over the telecom C-band. The ion near 1536.8 nm is found at 120 GHz detuning from the center of the inhomogeneous broadening (35 GHz) of the C\textsubscript{2} site, showing a full-width-at-half-maximum (FWHM) linewidth of 2.0 MHz. The g\textsuperscript{(2)}($\tau$) auto-correlation measurement on this peak confirms single photon emission with g\textsuperscript{(2)}(0) = 0.06 (right inset). The optical lifetime of this emitter in C\textsubscript{2} site was shortened from 8.5 ms to 0.14 ms (left inset), giving cavity QED parameters $\{g_0,\kappa,\gamma\} = 2\pi\times$\{1~{\rm MHz}, 3.4~{\rm GHz}, 18.7~{\rm Hz}\} and a Purcell enhancement of 60 fold which factored in a 22\% branching ratio (SI Section 2.2). Tuning the cavity to 1546.6 nm, an ion detuned 130 GHz from the center of the C$_{3i}$ site inhomogeneous distribution (46 GHz) showed a significantly narrower PLE linewidth of 0.19 MHz. Due to a weaker coupling strength (g$_0$ = 2$\pi\times$0.28 MHz) from a pure magnetic dipole allowed transition (Methods), the single Er emitter in the C\textsubscript{3i} site showed a modest lifetime reduction to $\mathrm{T_1}$ = 1.0 ms. Nevertheless, the stark contrast reveals an important role of symmetry on Er optical coherence in the two sites: while both C$_2$ and C$_{3i}$ Er emitters experience optical dephasing by coupling to a magnetic noise bath, the C$_2$ emitter can acquire additional broadening by coupling to electric noise due to its un-quenched permanent electric dipoles. Using our optical-microwave integrated qubit platform, correlated noise spectroscopy between the spin and optical transitions in each site can offer deeper insights for engineering the optical coherence of both Er emitters.

To elucidate optical spectral diffusion dynamics, we first focus on the C\textsubscript{2} site emitter. Exploiting in-situ tuning of the Purcell factor (SI Section 2.3, Fig.~S15), we swept the cavity-emitter detuning to reveal an evolution of the Er optical linewidth over a millisecond temporal window. Without an applied magnetic field, the optical linewidth shows a short-term broadening to $\approx$4 MHz as T\textsubscript{1} was tuned (red data in Fig.~\ref{fig4}b), likely due to slow spectral diffusion by paramagnetic impurities. Application of a modest B field ($\geq$250 mT) appeared to freeze the majority of the magnetic noise, leading to a reduction in linewidth to 0.95 MHz (blue data) that does not vary with emitter lifetimes from 200$\mu$s to 1.3 ms. The residual dephasing of $\sim$1.0 MHz can be accounted for using a model based on frozen paramagnetic spin baths \cite{erbiumSDpaper} (blue dashes in Fig.~\ref{fig4}b) (SI Section 2.4), which points to a fast ($\ll$200 $\mu$s) spectral diffusion process by non-magnetic noise, possibly related to fluctuating electric-dipoles from tunneling two-level-systems (TLS) that are evidently present in our device (SI Section 1.10). For assessing the long-term spectral stability, we also repeated single Er PLE scans over $\approx$10 hours (Fig.~\ref{fig4}c). The central peak of the PLE spectrum showed a room-mean-square (RMS) variation $\ll$1 MHz, attesting to the absence of a significant long-term drift.

Next, we turn to the C\textsubscript{3i} Er to probe an optical coherence limited only by the magnetic noise. In C\textsubscript{3i} site, the excited (g$_e \approx$0) and ground state (g$_g$=3.2) g-factors of Er result in a completely correlated spin-optical coherence. We measured spin spectral diffusion (top of Fig.~\ref{fig4}d) (SI Section 1.5) on the g=3.2 microwave transition at B=130 mT using three-pulse stimulated spin echoes and obtained a maximum spin linewidth of 2.9$\pm$0.5 kHz over a 10 ms scale. Subsequently, transient spectral holeburing \cite{Weiss:21} on the spin-preserving optical transition of the same site Er ions (bottom of Fig.~\ref{fig4}d) yielded a minimum optical Rabi linewidth (that is also the free-induction-decay linewidth) $\Gamma_{\rm {opt}}$ = 161 kHz, measured as spectral hole half-width at the lowest optical Rabi frequency $\mathrm{\Omega}$ = 2$\pi\times$33 kHz. This narrow linewidth was consistent with the single Er PLE full-width-half-maximum of 190$\pm$10 kHz at the same excitation power. Increasing optical power showed an expected broadening of $\Gamma_{\rm {opt}}$, which closely followed a transition from the Bloch limit \cite{PhysRevLett.50.1269} in the low power regime (grey dashes) to the Redfield limit \cite{PhysRevLett.79.2963} for high powers (grey dots). From the linewidth expression in the Bloch limit \cite{PhysRevLett.50.1269}

\begin{equation}
\mathrm{2\pi  \Gamma_{Opt} = \Big( \frac{1}{T_2^*}  
 +\sqrt{  \frac{1}{(T_2^*)^2} + \frac{\Omega^2 T_1}{T_2^*} } \Big)}
\end{equation}

\noindent and ensemble-averaged $T_1$ =3 ms, we calculate an optical $T_2^*$ = 122 $\pm$ 5 $\mu$s and a $T_2^*$ linewidth $\mathrm{\gamma}$ of $\mathrm{1/\pi T_2^*}$ = 2.6 kHz, in excellent agreement with the measured spin spectral diffusion linewidth, thus validating a correlated spin-optical magnetic-noise-only dephasing mechanism. Furthermore, the axial symmetry of the C\textsubscript{3i} site strongly forbids spin-flipping optical transitions, leading to a high cyclicity of $\sim$1000 that is desirable for high-fidelity spin readout \cite{reinemerthesis} (Methods).\newline

\noindent  {\bf Discussion and outlook}

\noindent  The kilohertz optical dephasing rate of Er spin qubits in C$_{3i}$ site is among the longest optical coherence reported for single Er dopants and is already on par with or surpassing the prevailing rare-earth doped bulk single crystals. Combined with a simultaneous millisecond spin coherence, we showcase a telecom-band spin-photon interface synthesized at a wafer-scale and fully fiber-integrated for plug-play deployment in a quantum network. While current device parameters (T$_2^*$/2T$_1\sim$0.1) do not reach unit cooperativity and fully transform-limited linewidth, improvement to the fiber cavity finesse ($\times$6) and mode alignment of atomic dipoles ($\times$3) will lead to a realistic cavity-QED regime of $\{g_0,\kappa,\gamma_h\} = 2\pi\times$\{1~{\rm MHz}, 1~{\rm GHz}, 1~{\rm kHz}\}, hence a high single-ion cooperativity and fully indistinguishable emission. For even stronger coupling, a slot-mode Y$_2$O$_3$ on silicon-on-insulator photonic resonators similar to \cite{Xu:20, Xu:22} can offer potentially another order-of-magnitude higher cooperativity without degrading Er optical coherence. Doing so will facilitate fast single-shot spin readout, enable $\mathrm{Er^{3+}}$-$\mathrm{Er^{3+}}$ remote entanglement, single Er-photon gate \cite{Reiserer2014} and non-destructive detection of telecom photons \cite{Obrien2016}. Beyond single qubits, the strong dipolar coupling between $\mathrm{Er^{3+}}$ pairs from either C\textsubscript{2}/C\textsubscript{3i} site or between dual sites \cite{Hu_2022} can realize decoherence-protected singlet-triplet qubits \cite{PhysRevLett.117.037203}, two-qubit entanglement, and logic gates. For the latter, a pair of $\mathrm{Er^{3+}}$ in $\mathrm{C_2}$ and $\mathrm{C_{3i}}$ sites combine the long coherence times of $\mathrm{C_{3i}}$ site with fast optical control and readout offered by $\mathrm{C_2}$ site.

Given the current spin coherence limit by the Er-Er dipolar interactions, we foresee ample room to enhance the Er qubit coherence times by lowering the doping concentrations to a part per millionth level. Isolating an $\mathrm{^{167}Er^{3+}}$ ion in $\mathrm{Y_2 O_3}$ and exploiting the available zero-first-order-Zeeman (ZEFOZ) transitions in this material \cite{y2o3esr2022} also provides a pathway towards ultra-long Er coherence. Following first demonstration of the epitaxial rare-earth qubit platform, we expect ensuing optimization and improvement on the thin film material quality: hybrid oxide growths combining MBE and chemical vapor deposition (CVD) \cite{DICVD} may produce a matrix with further reduced paramagnetic impurities or charge-trapping defects (e.g. F$^{+}$ centers (SI 1.6, 1.10)); co-doping Er with other rare-earths (e.g. Eu) that serve as charge-trapping centers \cite {Liao2023} could stabilize the local charge configuration and reduce optical spectral diffusion. Single $\mathrm{Er^{3+}}$, especially in C$_2$ site, in this regard, can serve as a sensor qubit\cite{Zhang2019} for mapping out the nanoscopic electric noise environment to guide material optimization. Finally, the thin film platform is amenable to integration with hybrid quantum degrees of freedom, such as coupling $\mathrm{Er^{3+}}$ to surface acoustic wave or suspended optomechanical resonators where strong confinement of phonons allows for fast spin manipulations as well as the study of novel spin-phonon physics. Lithographically patterned $\mathrm{Er^{3+}}$ doped films conforming to the overlapping modes of optical and microwave cavities can considerably boost microwave-to-optical quantum transduction efficiency \cite{PhysRevLett.113.203601, PhysRevA.100.033807}.
    
\begin{acknowledgments}
\noindent We thank David I. Schuster and Tanay Roy for their help on the ESR microwave experiment. We acknowledge Andrew Kamen, Xiaoyang Liu, Chao-Fan Wang, Noah Johnson for their technical assistance with the spectroscopy setup and device simulations. We thank Shuolong Yang, Qiang Gao, Woojoo Lee for setting up the oxide MBE facility at the University of Chicago, and Alexey Lyasota and Ian Berkman for the discussion on optical linewidth spectroscopy. We acknowledge MIT Lincoln Laboratory for providing the Josephson traveling-wave parametric amplifier (TWPA) and Argonne National Laboratory for the initial thin film growth. We thank Matthew Shaw for providing the single photon nanowire superconducting detectors. This work was funded by a National Science Foundation (NSF) Faculty Early Career Development Program (CAREER) award (grant no. 1944715) and Army Research Office (ARO) Young Investigator program award (W911NF2010296). Equipment funding was provided by the Department of Defense Office of Naval Research (N00014-22-1-2281) and the Department of Energy QNEXT quantum science and engineering research center. The work in Basel was funded by NCCR QSIT and by the Swiss National Science Foundation project $\mathrm{200020\_204069}$. 
\end{acknowledgments}

\clearpage

\noindent {\bf Methods}

\noindent {\bf $\mathbf{Y_2O_3}$ thin-film growth and integration}
Yttrium oxide ($\mathrm{Y_2O_3}$) has a cubic crystal structure ($T_h^7$ space group) with 16 formula units per unit cell \cite{harristhesis, reinemerthesis}. These 32 $\mathrm{Y^{3+}}$ sites can be grouped into two classes with 24 sites of $\mathrm{C_2}$ point group symmetry and 8 sites of $C_{3i}$ point group symmetry (Fig.~\ref{fig1}a), where each of the 32 yttrium ($\mathrm{Y^{3+}}$) ions in a unit cell can be substituted with an $\mathrm{Er^{3+}}$ ion with equal probability \cite{Sheller1} \cite{Schafer}.

Er$^{3+}$-doped $\mathrm{Y_2O_3}$ thin-films were grown epitaxially on silicon (111) wafers using molecular beam epitaxy \cite{doi:10.1063/1.5142611} technique thanks to a small lattice mismatch between $\mathrm{Y_2 O_3}$(222) and Si(111) plane. The MBE growth was performed at temperatures between 600 and 920 $^{\circ}$C and monitored in-situ during growth using reflection high electron diffraction (RHEED) as described in \cite{doi:10.1063/1.5142611}. The film thickness is 100 nm for the fiber-integrated qubit chip, and a thicker film of 1.5 $\mathrm{\mu m}$ on a 3 inch $\mathrm{Y_2 O_3}$ wafer was used for wafer-scale spectroscopy. All the films have the same doping profile with a 40 nm undoped buffer layer at the top surface and bottom $\mathrm{Y_2O_3}$/Si interface and were annealed in air at 600 $^{\circ}$C for 1 hour prior to device packaging. 

To integrate the $\mathrm{Y_2O_3}$ film onto a DBR mirror substrate, the thin-film sample and the DBR substrate are diced into 10$\times$10 mm chips and put in hot N-Methylpyrrolidone (NMP) for $>$ 5 hours for cleaning. The $\mathrm{Y_2O_3}$ chip is then bonded on the mirror using HSQ e-Beam resist as an adhesive layer by pressing them together using a home-built clamp and baked on a hot plate at 250 $^{\circ}$C for an hour. The samples are then removed from the clamp and annealed in a rapid thermal processing (RTP) tool at 600/500 $^{\circ}$C in air for 1 hour, which turns the HSQ to thermally stable oxide. Following a successful bonding, the back silicon handle is removed by reactive ion etching. A spacer layer of Si$_3$N$_4$ was deposited on the chip to planarize it and adjust the device thickness. 

\noindent
{\bf Superconducting microwave resonator and measurement setup} The co-planar superconducting resonator has a design impedance of 9.4 $\mathrm{\Omega}$, with simulated capacitance $\mathrm{\approx}$2.8-3.2 pF and inductance $\mathrm{L = 1/ 4 \pi^2 \omega^2 C} \approx$ 260 pH. The resonator frequency was tuned in 5.5-5.9 GHz by varying the width and spacing of the inter-digitated finger (IDF) capacitor in the range of 3.3 - 3.5 $\mathrm{\mu m}$. The stray inductance from the capacitor was $\approx$ 200 pH, and the inductor loop has a diameter 20 $\mathrm{\mu m}$ and an inductance $\approx$ 60 pH. For measurements only involving ESR, we adopt a flip-chip mounting technique for high-throughput device characterization \cite{PhysRevB.90.075112, PhysRevLett.105.140502, Staudt_2012, anisotropicg}. The resonators were first patterned on a high-resistivity silicon substrate following a recipe described in SI Section 1.1.  Microwave resonators were mounted on a copper PCB/holder, which were mounted on the cold finger of the 8 milliKelvin stage of a Bluefors LD-250 dilution refrigerator, and a 3-axis vector magnet (AMI 430) was used to apply a magnetic field in-plane of the resonator. The device used in Fig.~2 had a frequency of 5.81 GHz, intrinsic Q $\mathrm{\approx}$ 370,000, and external Q $\mathrm{\approx}$ 3,000. The three resonators in Fig.~3 had frequencies of 5.55, 5.58, and 5.83 GHz with intrinsic Q of $\mathrm{\approx 12,000}$ with an external Q of $\mathrm{\approx}$ 1484, 2666, 1620, respectively (SI Table 1). The resonators showed varying degrees of asymmetric lineshape as seen in Fig.~\ref{fig1}{\bf c} due to reflections from imperfect impedance matching in the transmission line  \cite{doi:10.1063/1.4817512}. Continuous-wave ESR measurements were performed using a vector network analyzer with -80 dBm on-chip power. The pulsed electron spin resonance (ESR) measurements used a low noise amplifier chain comprised of a Josephson traveling wave parametric amplifier (TWPA) \cite{twpadoi:10.1126/science.aaa8525} at the 8 mK stage, a low noise amplifier at 4 Kelvin and room temperature electronics for high sensitivity spin echo detection. Complete details of the setup are included in SI Section 1.1 and Fig.~S1.

\noindent
{\bf Cryogenic tunable fiber cavity and optical measurement setup}
The dimpled fiber mirror was prepared by creating a concave depression using CO\textsubscript{2} laser ablation as described in \cite{10.1063/1.3679721}. Following the convention in ~\cite{10.1063/1.3679721}, our fiber dimple had a radius of curvature R of 40 $\mathrm{\mu}$m, a depth t of 1.54 $\mu$m and structure diameter d of 10.35 $\mathrm{\mu}$m. The fiber dimple was subsequently coated with the same DBR mirror stack as on the substrate, which has a nominal transmission of 150 ppm. The theoretical cavity finesse is 21,000. 

The fiber mirror is mounted in a fiber chuck and secured with thin copper shims. The fiber tip and the device chip are brought into a rigid contact using a piezoelectric nanopositioner to form a stable, small (L=$\mathrm{3/2}\lambda$) Fabry-Perot cavity without active stabilization. The fiber mirror showed a one-way reflection of $\approx$80 \%, which included transmission through the mirror coating and loss due to a mismatch between the cavity and the fiber mode. Even after the rigid contact, the optical cavity resonance is still coarsely tunable up to 40 nm in wavelengths at cryogenic temperaturesby stepping the nanopositioner. At the same time, continuous fine-tuning at a rate of 1.3 GHz/V over 80 GHz is achieved by adjusting the piezo voltage. The measured finesse of an empty cavity (no Y$_2$O$_3$ layer) ranges from 15,000 to 20,000 over the tunable wavelengths. Optical measurements on the C$_2$ site were performed on the cavity with a finesse of 19,300 (Q = 58,000), while on the C$_{3i}$ site, the cavity finesse was 16,000 (Q = 48,000). Furthermore, from the change of cavity finesse with and without $\mathrm{Y_2 O_3}$ film, we calculate an upper bound on the optical absorption of the $\mathrm{Y_2O_3}$ thin film to be 1.8 dB/m, indicating an outstanding optical quality, which bodes well for further photonic integrations. The fundamental cavity mode splits into two orthogonal linear polarization modes separated by 30 GHz, which has been previously reported in fiber cavities \cite{Tomm2021, Najer2019}. A slight asymmetry of the fiber cavity reflection spectrum in Fig.~1c is attributed to the Fano effect and can be minimized by adjusting the polarization of the excitation laser.

The optical setup comprises of three acousto-optic modulators (AOM) in series for generating pulses with a high extinction ratio of 160 dB and an electro-optic modulator (EOM) for a fine frequency sweep of the laser. Photoluminescence from Er$^{3+}$ was detected by a superconducting nanowire single photon detector(SNSPD) with $\approx$70\% detection efficiency and a 2.1 count per second (cps) dark count rate. The laser was locked to a UHV stable reference cavity with short-term laser linewidth of 0.4 kHz and a long-term drift \textless 100kHz/day. The laser power on-chip is actively stabilized via variable optical attenuators.

\noindent
{\bf Selection rules for the Er\textsuperscript{3+} optical transition in C\textsubscript{3i} site}
Optical transitions of rare-earth dopants in sites with an inversion symmetry are magnetic dipole allowed only and are governed by the selection rule: $\Delta$J$\leq$ 1 and $\Delta$M\textsubscript{J}$\leq$ 1 \cite{reinemerthesis}. For C$_{3i}$ site Er$^{3+}$ optical transitions, the wavefunctions of the lowest crystal-field doublet Z\textsubscript{1} of $^{4}I_{15/2}$ and Y\textsubscript{1} of $^{4}I_{13/2}$ are dominated by M\textsubscript{J} = $\pm$15/2 and M\textsubscript{J} = $\pm$13/2 components, respectively. Therefore, spin-flipping optical transitions between Kramer double groups of Z\textsubscript{1} and Y\textsubscript{1} are largely forbidden due to a large change in $\Delta$ M\textsubscript{J} $\gg$1. This enforced selection rule was supported by an absence of spin-flipping transitions in the optical spectra of Er$^{3+}$ in $\mathrm{Y_2O_3}$ C\textsubscript{3i} site \cite{reinemerthesis}. In addition, we observed no change of photoluminescence after 1000 repetitions of pulsed excitation on the C\textsubscript{3i}, providing further evidence of a high cyclicity and negligible branching ratio for the spin-flipping transition. The magnetic dipole allowed optical transitions in the C$_{3i}$ site support two possible polarizations: either a linear polarization parallel to the C$_{3i}$ axis (normal to the device plane and longitudinal to the fiber cavity) or a circular polarization perpendicular to the C$_{3i}$ axis. From the maximum Purcell enhancement we observed from cavity-emitters in the C$_{3i}$ site, we deduce that their photon emission is most likely circularly polarized. 

 \bibliography{Erqubit}
 
 \clearpage

\newcommand{\beginsupplement}{%
        \setcounter{table}{0}
        \renewcommand{\thetable}{S\arabic{table}}%
        \setcounter{figure}{0}
        \renewcommand{\thefigure}{S\arabic{figure}}%
        \renewcommand{\thesection}{S\arabic{section}}
        \renewcommand{\thesubsection}{S\arabic{section}}
     }

\onecolumngrid

      \beginsupplement

\begin{center}
{\bf Dual epitaxial telecom spin-photon interfaces with correlated long-lived coherence: Supplementary Information}
\end{center}

\raggedbottom

\tableofcontents


\section{Spin coherence spectroscopy}
\renewcommand{\thesubsection}{1.\arabic{subsection}}
\noindent All the electron spin resonance (ESR) measurements mentioned here except in section \ref{inhomcpmgsection} and \ref{TLSsection} have been performed on the sample mentioned in Fig.~2 in the main text.

\subsection{ESR measurement setup and device fabrication}

\noindent
{\bf Superconducting resonator fabrication and flip-chip mounting}
The resonators were fabricated on a Y$_2$O$_3$ wafer or a high-resistivity silicon substrate. The substrate was solvent-cleaned and then transferred to the e-beam deposition chamber and baked in a vacuum (\textless 1e-7 mBar) at 300 $^\circ$C for 1 hour, followed by an overnight vacuum pump. A 75 nm niobium layer was subsequently deposited with e-beam evaporation, followed by optical lithography on a positive photoresist mask (Az 703 MIR) and reactive-ion etching (RIE) with $\mathrm{CF_4}$, $\mathrm{CHF_3}$ and Ar plasma for 4 min 30 sec for etching a 75 nm niobium layer.

A 100 nm niobium spacer layer was further patterned on the resonator chips using a liftoff process (Az nLof2020 photoresist) in order to protect the resonator from damage during the flip-chip mounting of the $\mathrm{Er^{3+}}$ doped thin-film samples. The device was dipped in N-Methyl pyrrolidone (NMP) at 80 $^\circ$C for 10 hours to remove the photoresist and finally cleaned with Acetone, IPA, and DI water.

For ESR-only measurements, the Er:Y$_2$O$_3$ chip was flip-chip mounted with a superconducting device chip and held in place with vacuum grease (Apiezon N). The triangular end of the sample was localized over the inductor wire in order to avoid dielectric losses from a strong electric field above the interdigitated finger (IDF) capacitors.\newline

\noindent
{\bf Continuous-wave ESR} CW ESR was performed by measuring the resonator transmission spectrum $\mathrm{S_{21}}$ as a function of an in-plane magnetic field with a vector network analyzer (Keysight E5071B). The estimated on-chip power was $\mathrm{\approx}$ -80 dBm, and the magnetic field was ramped in steps of 0.1 mT. The resonator spectrum was fitted to the model for LC superconducting resonators coupled to a transmission line from ~\cite{doi:10.1063/1.4817512} Eq.~(23) to obtain the center frequency $\mathrm{\omega_0}$, quality factors $\mathrm{Q_i}$, $\mathrm{Q_e}$, and $\mathrm{Q_{\alpha}}$,

\begin{equation}
    S_{21} = \frac{1+ 2jQ_i(\frac{\omega - \omega_0}{\omega_0})}{1+ \frac{Q_i}{Q_e} + j\frac{Q_i}{Q_{\alpha}} + 2j Q_i(\frac{\omega - \omega_0}{\omega_0})} 
\end{equation}

\noindent
where $\mathrm{\omega_0}$ is the resonator frequency, $\mathrm{Q_i}$ is the intrinsic quality factor, $\mathrm{Q_e}$ is the external quality factor, and $\mathrm{Q_{\alpha}}$ is introduced to account for the asymmetry of the transmission spectrum \cite{doi:10.1063/1.4817512}.
The fitted intrinsic linewidth $\mathrm{k_i}$ and frequency $\mathrm{\omega}$ was plotted against the B field, and a slow-varying background was subtracted to obtain the spectrum in Fig.~2(b) in the main text.\newline

\begin{figure*}[]
    \centering
    \includegraphics[width=500pt]{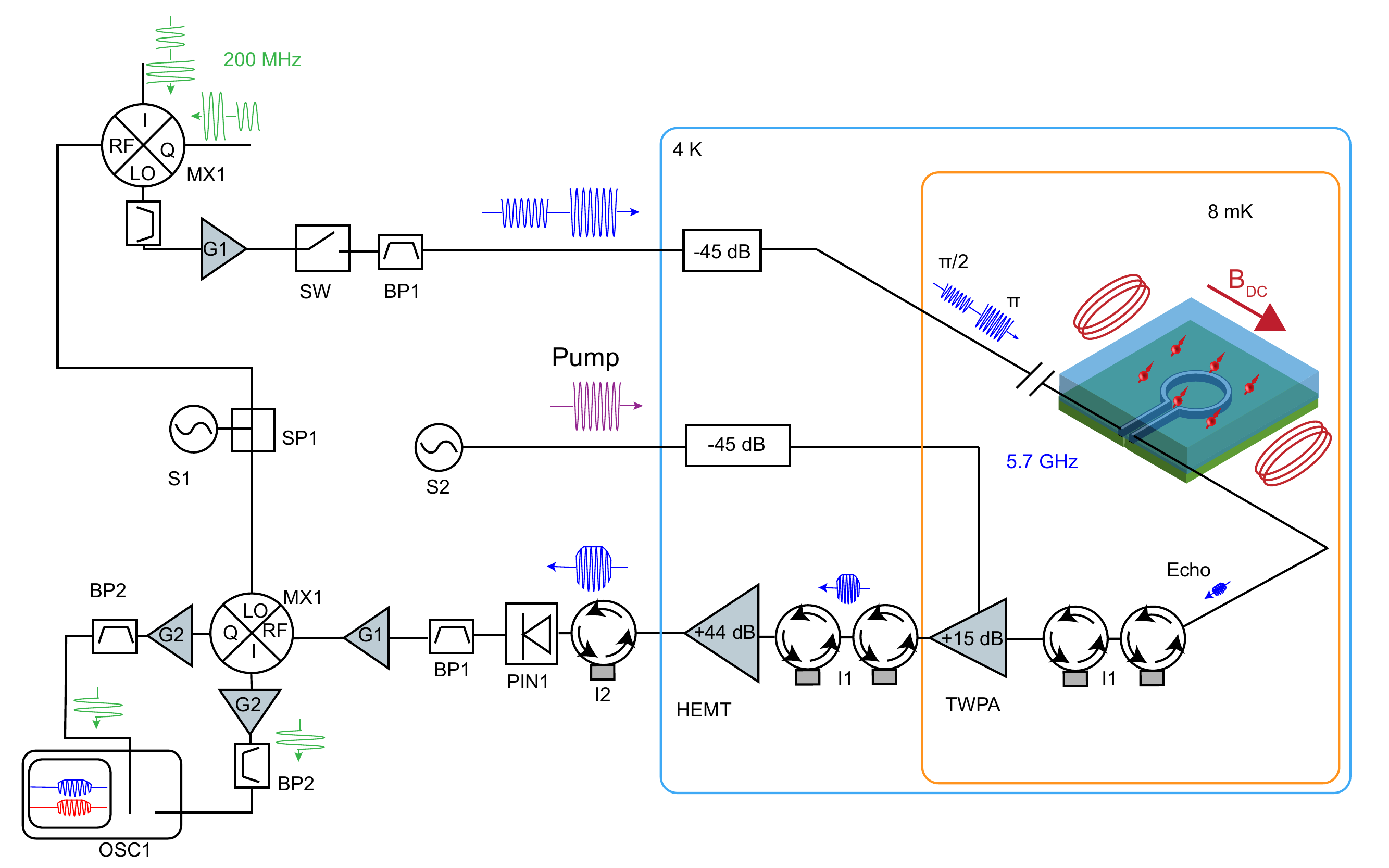}
    \caption[Pulsed ESR setup]{Pulsed ESR setup. The spin control pulses were generated using IQ upconversion, and the spin echo was amplified by a chain of cryogenic and room temperature amplifiers followed by downconversion to read the time trace on a fast oscilloscope. }
    \label{fig:esrsetup}
\end{figure*}

\noindent
{\bf Pulsed ESR} The pulsed spin echo measurement setup is shown in Fig.~\ref{fig:esrsetup}. An arbitrary waveform generator (Zurich Instrument HDAWG) was used to generate 200 ns long I and Q pulses at 200 MHz frequency and upconverted to 5.7 GHz using an IQ mixer (MX1: Marki MMIQ 0218LXPC) with the estimated on-chip $\mathrm{\pi}$ pulse power of -65 dBm (300 pW) at the sample after amplification (G1:Mini-Circuits ZX60-83-LNS+) and 45 dB cryogenic attenuation. A pulsed pump generated using IQ upconversion (S2: Signal Core SC5511A and Marki MMIQ 0307LXP) was sent to the TWPA during spin echo acquisition time to turn on the gain of the TWPA. The TWPA pump frequency and power were tuned to obtain a maximum signal-to-noise (SNR) gain of 10 dB while avoiding spurious pump tones interfering with the spin echo frequency. The spin echo signal was amplified by TWPA (+15 dB) at the 8 mK stage, followed by HEMT (Low noise factory $\mathrm{LNF-LNC4\_8F}$ +44 dB) at the 4K stage. Dual Junction cryogenic isolators (I1: Low noise factory $\mathrm{LNF-CICIC4\_8A}$) were placed in between the resonator and TWPA and the TPWA and HEMT to prevent signal reflection and thermal noise leakage to the 8 mK stage. A diode limiter (PIN1: Pasternack PE8023) and an isolator (I2: Fairview Microwave SFC0206S) were added to protect the room temperature readout electronics during pulses. Bandpass filters with 4.9-6 GHz passband (BP1: Mini-Circuits VBFZ-5500-S+) were placed in the pulse generation and readout chain to avoid amplifier saturation. The spin echo was further amplified with room temperature amplifier (G1:ZX60-83-LNS+,+21 dB) followed by demodulation (Marki MMIQ0218LXPC mixer) to obtain I$^\prime$, Q$^\prime$ signals at 200 MHz. The local oscillator (LO) signal to the demodulation mixer was supplied by the same source used for pulse generation (S1:Keysight N5181A). The demodulated spin echo was further amplified by an amplifier (Mini-Circuits ZX-60-P103LN+, +23 dB) and bandpass filtered (BP2: Mini-Circuits ZX75BP-204-S+) to obtain I$^\prime$, Q$^\prime$ traces on a fast oscilloscope (OSC1:Tektronix TDS5104). We note that the I$^\prime$, Q$^\prime$ signals obtained after demodulation have a phase offset with respect to the input I and Q signal used in pulses. Nevertheless, we measure only the amplitude of the spin echo signal, and therefore, we ignore the phase offset.

\begin{figure*}[t!]
    \centering
    \includegraphics[width=1.0\textwidth]{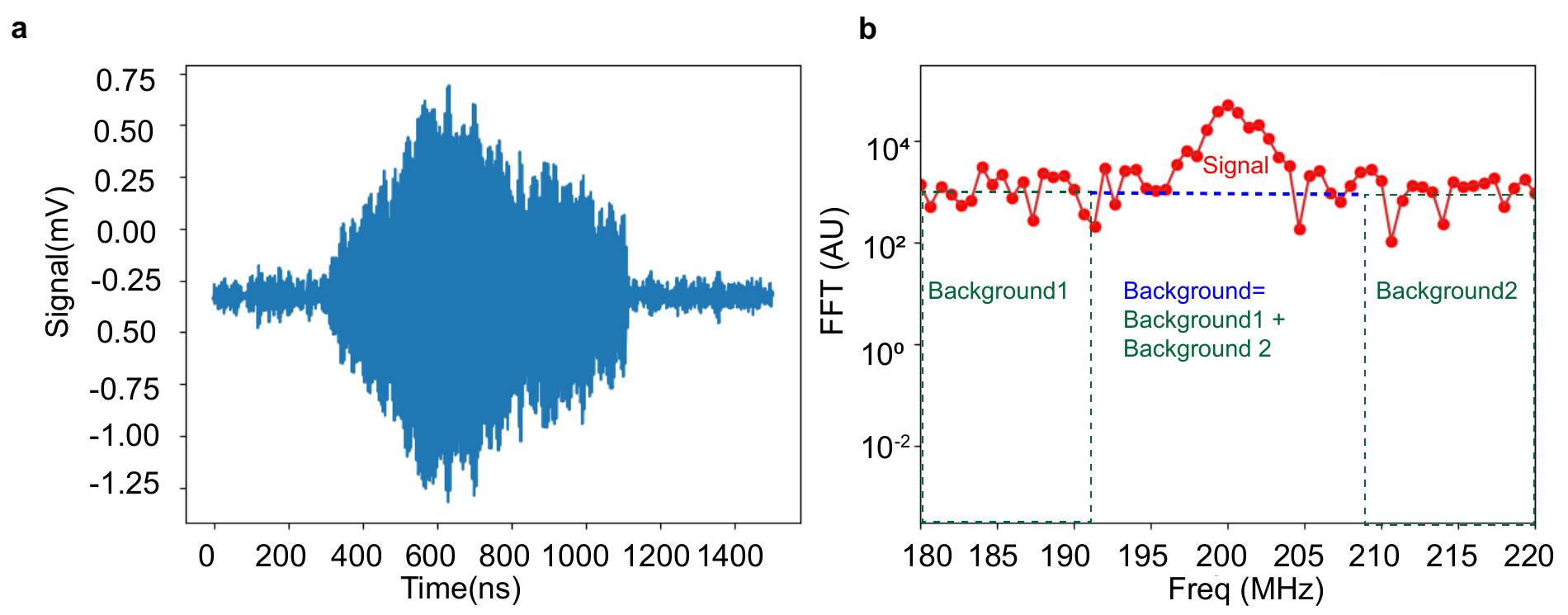}
    \caption[Spin echo processing]{ Processing of the spin echo signal. {\bf a} Time domain spin echo signal (I$^\prime$ quadrature). {\bf b} Fast Fourier transform (FFT) of the signal over 1500 ns window. The signal is integrated over a 20 MHz window, and the noise background (blue) is estimated by integrating the FFT in two 10 MHz windows to the left and right of the signal integration area (green). The total noise background(blue dash) can be estimated as a sum of the area in green dashes (background 1 and background 2) and is subtracted from the signal to get the background-subtracted signal for each I$^\prime$ and Q$^\prime$ quadrature.}
    \label{fig:fft}
\end{figure*}

 The time domain spin echo signal I$^\prime$(t) (Fig.~\ref{fig:fft}{\bf a}) was Fourier-transformed over an integration window of 1500 ns. Fast Fourier transform (FFT) spectrum of the signal I$^\prime$($\mathrm{\omega}$) in Fig.~\ref{fig:fft}{\bf b} shows a spin echo peak at 200 MHz sitting on a noise background pedestal (blue dashes). An integrated over bins in a bandwidth of 20 MHz around 200 MHz was performed in the FFT spectrum, and the area under the background noise pedestal (blue dashes) was subtracted to obtain the spin echo area (between the blue dash and red line). The background noise pedestal (Fig.~\ref{fig:fft}{\bf b} blue dashes) was estimated by summing the integral over two adjacent 10 MHz bandwidth bins (green dashes) around the signal integration window. Finally, the background subtracted spin echo signal for each channel I$^\prime$ and Q$^\prime$ was added in quadrature to obtain the total spin echo area. 
 
\subsection{Simulated and measured coupling of spins to microwave resonator}
\begin{figure*}
    \centering
    \includegraphics[width=0.8\textwidth]{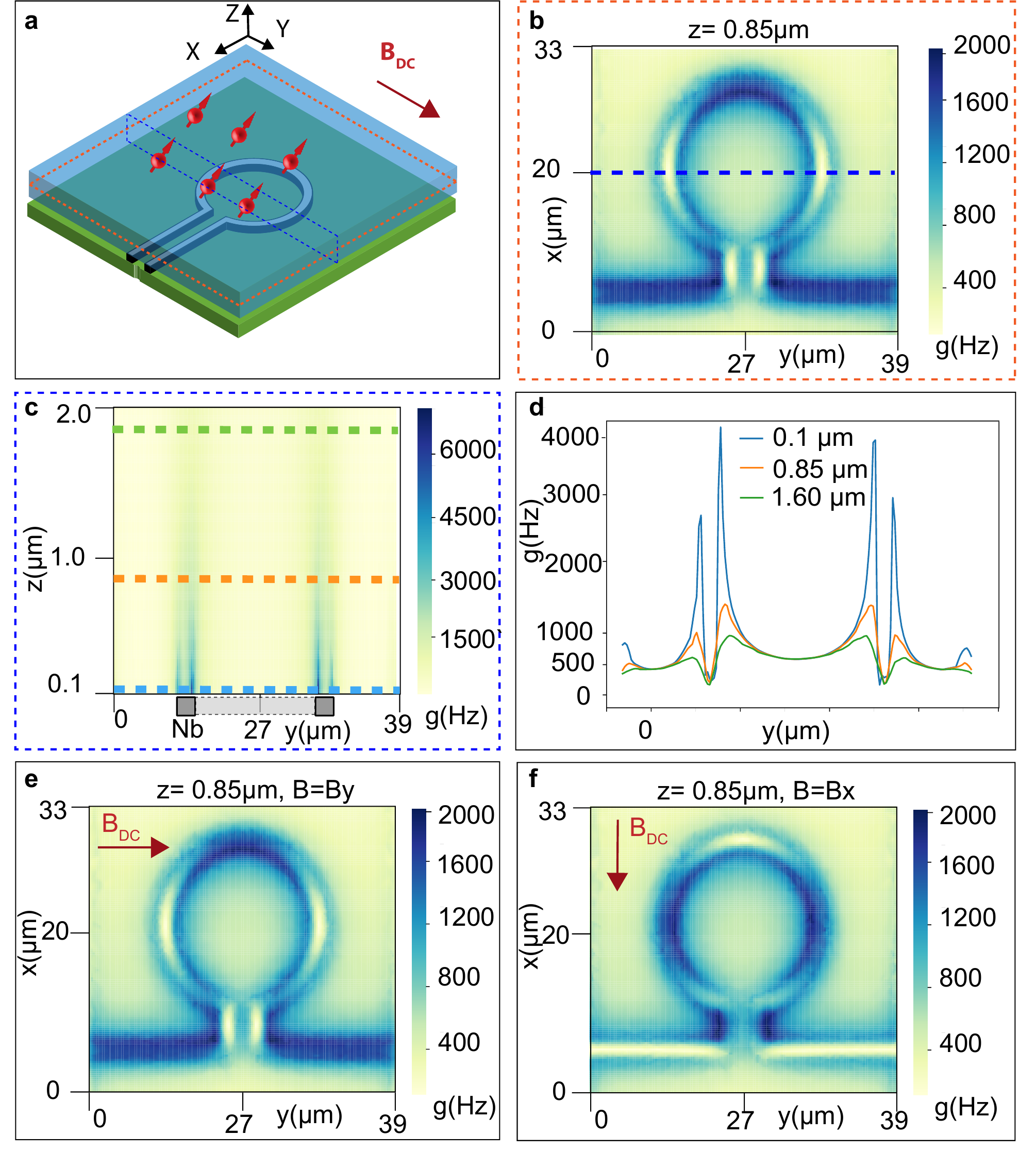}

    \caption[Simulated spin-resonator coupling]{Simulated single spin coupling g (Hz) above the microwave resonator for spins with $\mathrm{g_{AC}=7}$. {\bf a} Device schematic with the $\mathrm{Er^{3+}}$ doped $\mathrm{Y_2 O_3}$ sample mounted on the resonator in a flip-chip configuration. {\bf b} Coupling strength in the plane z=0.83 $\mathrm{\mu m}$ above the resonator. {\bf c} Coupling along a cross-section taken in the plane x=20 $\mathrm{\mu m}$. {\bf d} Three line cuts from (c) show a decrease in amplitude and an increase in spatial homogeneity of g  over the inductor loop with distance z. {\bf e} Coupling strength g for DC B field applied along Y and, {\bf f} X  direction. Different parts of the inductor contribute to spin driving depending on the direction of the applied B field.}
    \label{fig:resonatorsimulationl}
\end{figure*}

\begin{center}
\begin{table}
\begin{tabular}{||c c c c c||} 
 \hline
 Device & $\mathrm{\omega}$ (simulated) & $\mathrm{\omega}$ (measured) & $Q_{i}$  & $Q_{e}$\\ [0.5ex] 
 \hline\hline
 1 & 5.55 GHz & 5.81 GHz & 370043 & 3001 \\ 
 \hline
 2 & 5.55 GHz & 5.88 GHz & 11776 & 1484\\
 \hline
 3 & 5.31 GHz & 5.58 GHz & 13650 & 2666 \\
 \hline
 4 & 5.23 GHz & 5.52 GHz & 11054 & 1620\\ [1ex] 
 \hline
\end{tabular}
\caption{Simulated and measured frequency and quality factor of microwave resonators. Device 1 was used for results in Fig.~2 in the main text, and Device 2,3,4 for Fig.~3 in the main text.}
\label{fitQ}
\end{table}
\end{center}

\noindent
{\bf Resonator simulations} The spatial and spectral mode profile of the low-impedance resonators used in ESR measurements were simulated numerically with ANSYS\textregistered HFSS \cite{ansys}. Table.~\ref{fitQ} shows the simulated and measured frequency and quality factors for the resonators used in the ESR measurements in the main text. All measurements were performed at 8 mK and an on-chip power of $\approx$ -80 dBm. Device 1 in the table corresponds to the device used in Fig.~2 in the main text, and Device 2,3,4 were used in the measurements in Fig.~3{\bf b} in the main text. The simulated and measured frequencies show an offset of $\mathrm{\approx}$ 300 MHz, which could be accounted for by a slight deviation in the dielectric constant of the $\mathrm{Y_2 O_3}$ thin films. Nevertheless, the relative spacing of individual resonators is consistent between the two.

Figure~\ref{fig:resonatorsimulationl} shows the simulated single spin coupling g (Hz) in the plane above the resonator for spins with transverse g factor $\mathrm{g_{AC}}$ = 7 and magnetic field applied along the Y direction. Figure~\ref{fig:resonatorsimulationl}{\bf a} shows the schematics of the resonator and the magnetic field configuration. Figure~\ref{fig:resonatorsimulationl}{\bf b} shows the spatial profile of g at a distance z of 0.85 $\mathrm{\mu m}$ above the resonator with strong fringing fields along the omega inductor leading to increased single spin coupling. Fig.~\ref{fig:resonatorsimulationl}{\bf c} and Fig.~\ref{fig:resonatorsimulationl}{\bf d} show a cross-section and a line cut, respectively, taken in the plane x= 20 $\mathrm{\mu m}$. While the single spin coupling shows a decrease from the peak value of 4 kHz at z=100 nm to $\approx$ 800 Hz at z=1.6 $\mathrm{\mu m}$, there is greater spatial uniformity of g at z=1.6 $\mathrm{\mu m}$, which could be advantageous in uniform driving of the spin ensemble. The average spin coupling strength $\mathrm{g_{avg}}=g_{ens}/\sqrt{N}$ was calculated by averaging g over z= 0.1 to 2 $\mathrm{\mu m}$ and the cross-sectional area x= 0 to 30 $\mathrm{\mu m}$, y= 0 to 39 $\mathrm{\mu m}$.

We simulate a direction dependence of the spin coupling profile over the inductor as the DC magnetic field is swept in the XY plane. While the Z component of the resonator AC magnetic field is always perpendicular to the applied DC B field in the XY plane, as the DC field is swept along the XY plane, the segments of the inductor wire which are parallel to the DC B field additionally couple to the spins through the XY component of the resonator AC magnetic field. This leads to a variation in the spatial profile of single spin coupling strength (g) as well as total ensemble coupling $\mathrm{g_{ens} = \sqrt{\sum_i g_i^2 }}$. Fig.~\ref{fig:resonatorsimulationl}Fig.~\ref{fig:resonatorsimulationl}{\bf e} shows when the DC B field is applied along the Y direction, the horizontal sections of the inductor loop, as well as the straight segment of the wire on the base of the inductor loop, show stronger coupling strength g (dark blue regions). As the B field direction is changed to X (Fig.~\ref{fig:resonatorsimulationl}Fig.~\ref{fig:resonatorsimulationl}{\bf f}), the vertical components of the inductor wire now have a stronger spin coupling g. We observe a slightly higher average spin coupling $\mathrm{g_{avg}}$ of 843 Hz when the B field is applied along y (Fig.~\ref{fig:resonatorsimulationl}{\bf e}) as compared to 795 Hz when the B field is applied along X (Fig.~\ref{fig:resonatorsimulationl}{\bf f}) due to an additional contribution from the long horizontal straight section of the inductor wire. 

We did not observe a strong frequency dependence of Er spin $\mathrm{T_1}$ (see Section.~\ref{sectspinT1}). Thus, we can constrain the Purcell-enhanced spin relaxation time $\mathrm{T_1}$ to $>$5 seconds, which puts the upper limit on average single spin coupling g \textless 128 Hz (see Section.~\ref{sectspinT1}). This indicates the spacing between the sample and resonator is significantly larger than 100 nm (niobium spacer layer thickness) as we expect $\mathrm{g} \approx$ 800 Hz if the sample is at 100 nm distance from the resonator. This could be caused by imperfect sample mounting leading to an angle offset as well a vertical gap between the plane of the resonator and the $\mathrm{Y_2 O_3}$ sample. We indeed observed an angle offset up to 2.8$^{\circ}$ between the plane of the resonator and the sample in the CW-ESR angle scan (Fig.~2{\bf a} in the main text) and directly measured air gaps up to 2 $\mathrm{\mu m}$ between the sample and the resonator using SEM. This projects a further possible increase in ensemble spin coupling by optimization of the mounting process to reduce the gap between the resonator and the sample, which would allow for strong Purcell-enhanced spin relaxation through the resonator.\newline

\begin{figure*}
    \centering
    \includegraphics[width=0.99\textwidth]{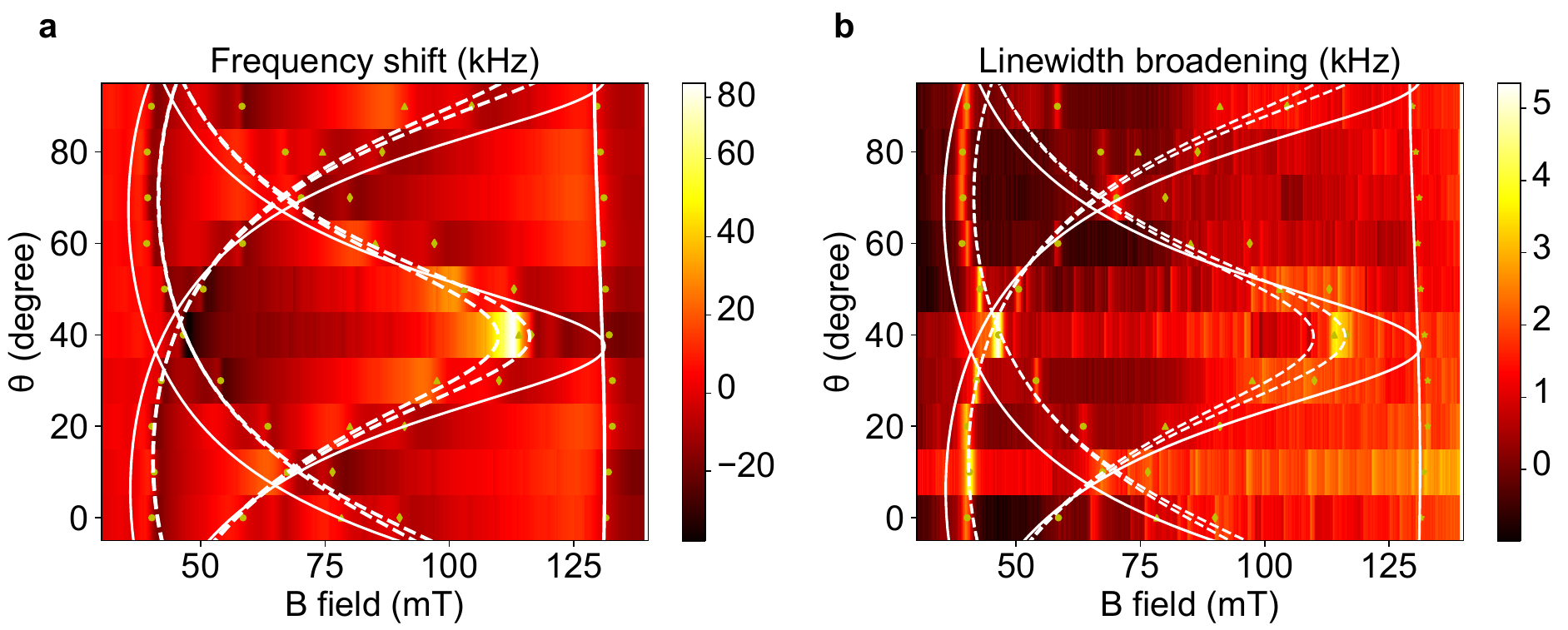}
    \caption[Dispersive and absorptive spin-resonator coupling]{Dispersive and absorptive coupling of $\mathrm{Er^{3+}}$ spins in $\mathrm{Y_2O_3}$ to superconducting resonators. {\bf a} Resonator frequency shows dispersive shifts ($\mathrm{\Delta \omega}$) when on-resonant with spins, resulting in a characteristic increase of resonator frequency followed by a decrease as the magnetic field is swept across the spin resonance fields. A strong dispersive signal can be clearly resolved for both high and low field transitions represented by a change in color from blue(dark blue) to red(light blue) in the heatmap. {\bf b} The absorptive coupling results in the broadening of resonator intrinsic linewidth ($\mathrm{\Delta k_i}$) when on resonant with spins. Low-field transitions show up with greater contrast in the absorptive scan. White dotted lines show predicted transitions based on simulations, and yellow dots show the center of the measured transition for both {\bf a} and {\bf b}.}
     \label{dispersiveabsorptive}
\end{figure*}

\noindent
{\bf Absorptive and dispersive ESR coupling} 
Absorptive and dispersive coupling to spins results in linewidth broadening and frequency shift of the resonator, as described in the main text. We observed dispersive as well as absorptive coupling of $\mathrm{Er^{3+}}$ spins to superconducting resonator as the magnetic field (B) is scanned in the plane of the resonator at an angle $\mathrm{\theta}$ (Fig.~\ref{dispersiveabsorptive}). Yellow dots show the center of the measured transition for both Fig.~\ref{dispersiveabsorptive}{\bf a} and Fig.~\ref{dispersiveabsorptive}{\bf b}.

The magnetic field was aligned to the plane of the superconducting resonator for each $\mathrm{\theta}$ by finding an optimum ratio of $\mathrm{B_{XY}}$ and $\mathrm{B_Z}$ for each $\mathrm{\theta}$. The optimum ratio of the XY and Z fields is found by maximizing the resonator frequency by tuning the Z field, and the field is then ramped up iteratively in the optimum direction \cite{doi:10.1063/1.5129032}. The alignment procedure was performed at 4 Kelvin in order to allow for quick thermal cycling above the niobium critical temperature of 9 Kelvin to cancel the magnetic vortices.
The white dashes in Fig.~\ref{dispersiveabsorptive}{\bf a} and ~\ref{dispersiveabsorptive}{\bf b} show predicted transitions based on the g tensor of $\mathrm{Er^{{3+}}}$ in $\mathrm{Y_2 O_3}$ for an applied magnetic field at an in-plane angle $\mathrm{\theta}$. To take into account a relative tilt between the (111) plane of the thin film and the XY plane of the resonator, we used an analytical expression for the tilt angle $\mathrm{\delta \phi_{offs}}$ varying between $\mathrm{\phi_{offs1}}=$ 2.8$^{\circ}$ and $\mathrm{\delta\phi_{offs2}}=$ 0.2$^{\circ}$ as a function of $\mathrm{\theta}$ following the analytical expression, $
\mathrm{\delta \phi_{offs}(\theta) = tan^{-1} ( tan(\delta \phi_{offs1})cos(\theta+\theta_0)}  
\mathrm{+ tan(\delta \phi_{offs2})sin(\theta+\theta_0)) }$
where $\mathrm{\theta_0}$ = 30 degrees was an in-plane angle offset between the X axis of the magnet and direction [2,1,-1] in the plane of the $\mathrm{Y_2 O_3}$ thin film. The above expression for a variable tilt angle $\mathrm{\phi_{offs}}$ was taken into account in simulations to generate the predicted transitions from  $\mathrm{C_{2}}$ (white dashes) and $\mathrm{C_{3i}}$ (white solid lines) sites as a function of angle for Fig.~\ref{dispersiveabsorptive}{\bf a} and Fig.~\ref{dispersiveabsorptive}{\bf b}. While all the transitions from $\mathrm{C_2}$ site are clearly resolved in the angle scans, only one out of the 4 $\mathrm{C_{3i}}$ sub-sites with symmetry axis along [1,1,1] direction could be observed. This is explained due to a large overlap of transition from the $\mathrm{C_{3i}}$ subsites with symmetry axis along [1,-1,1], [-1,1,1] and [1,1,-1] with stronger transitions from the $\mathrm{C_{2}}$ sub-sites. 

Figure~\ref{dispersiveabsorptive}{\bf a} shows the dispersive resonator frequency shifts as a function of B field for different angles ($\mathrm{\theta}$). The resonator frequency ($\mathrm{\omega}$) after field alignment shows a quadratic decrease as a function of the B field due to an increase in kinetic inductance ($\mathrm{\omega = \omega_0 - cB^2}$) \cite{patrickthesis}. This quadratic dependence on the resonator frequency shifts is subtracted to obtain the frequency shifts $\mathrm{\Delta \omega}$ plotted in Fig.~\ref{dispersiveabsorptive}{\bf a}. We observed dispersive frequency shifts for both high and low-field transitions, as shown by the regions with a change in color from black to yellow.

Figure~\ref{dispersiveabsorptive}{\bf b} shows the increase in resonator linewidth due to absorptive coupling to spins. The intrinsic quality factor ($\mathrm{Q_i}$) was obtained by fitting the model in Eq.~(1) to the resonator spectrum for each B field. The intrinsic linewidth $\mathrm{k_i= \omega/Q_i}$ was subtracted with the average intrinsic linewidth $\mathrm{k_{i,avg}}$ for each scan $\mathrm{\theta}$ to obtain the linewidth broadening $\mathrm{\Delta k_i = k_i - k_{i,avg}}$ plotted in Fig.~\ref{dispersiveabsorptive}{\bf b}. The strongest high field transition for $\mathrm{C_2}$ g=3.4 at 113-117 mT ($\mathrm{\theta=}$ 40 degree) as well as low field transitions are clearly resolved in both absorptive and dispersive signals. However, an increase in $\mathrm{k_i}$ for a high B field leads to a background to the broadening signal, and some of the high field transitions could not be clearly resolved in the absorptive broadening signal.\newline

\noindent
{\bf Estimate of $\mathbf{Er^{3+}}$ density} We use the dispersive coupling signal from Fig.~\ref{dispersiveabsorptive}{\bf a} to estimate erbium concentration. The maximum ensemble coupling was $\mathrm{g_{ens}}$ of 3.07(0.02) MHz for the transition from $\mathrm{C_{3i}}$ sub-site at B= 130 mT, $\mathrm{\theta}$ = 10 degree, which has a $g_{AC} = 12$ and single spin coupling $\mathrm{g}\sim$ 128 Hz (see Section.~\ref{t1vsofftemp}). This gives a total spin population N $\sim$ 574.7 $\times 10^{6}$ spins. Dividing by the sample volume of 1256 $\mathrm{\mu m^2}$ $\times$ 1.4 $\mathrm{\mu m}$ we get a density $\mathrm{n_{Er,C_{3i}}}$ of 3.29 $\times 10^{17}$ spins/$\mathrm{cm^3}$ or 12.4 ppm for $\mathrm{Er^{3+}}$ in $\mathrm{C_{3i}}$ site.

 Similarly we obtained a maximum ensemble coupling $\mathrm{g_{ens}}$ = 1.88(0.029) MHz for the transition from $\mathrm{C_2}$ sub-sites at B= 113 mT, $\mathrm{\theta}$ = 40 degree, corresponding to a total number of spins N $\sim$ 636.0(0.01) $\times 10^6$ spins assuming $\mathrm{g}$= 74.5 Hz ($\mathrm{g}$ 128 Hz for $\mathrm{C_{3i}}$ transition with $\mathrm{g_{AC}} =12$, therefore for $\mathrm{C_2}$ transition with $\mathrm{g_{AC} =7}$ we get $\mathrm{g}$ 128(7/12) Hz). Dividing by the sample volume of 1256 $\mathrm{\mu m^2}$ $\times$ 1.4 $\mathrm{\mu m}$ we get a density $\mathrm{n_{Er,C_2}}$ of 3.61 $\times 10^{17}$  spins/$\mathrm{cm^3}$ or 13.6 ppm ($\mathrm{n_{Y} = 2.66 \times 10^{22} spins/cm^3}$) for $\mathrm{Er^{3+}}$ in $\mathrm{C_{2}}$ site. Multiplying by a factor of 6 (six $\mathrm{C_{2}}$ sub-site orientations), we get a total density of 81.4 ppm for all $\mathrm{Er^{3+}}$ in $\mathrm{C_{2}}$ sites. \newline

\noindent
{\bf Minimum number of spin sensitivity ($\mathrm{N_{spin}^{min}}$}) We use the CPMG measurement on the 130 mT transition from $\mathrm{C_{3i}}$ site in Fig.~2 of the main text to estimate the minimum number of spins detectable. For the CPMG spin echo at time = 82 $\mathrm{\mu s}$, we measure a signal with SNR = 60 for $\mathrm{24\ \times 10^6}$ spins being addressed, which leads to a minimum sensitivity of 36 $\mathrm{\times  10^4}$ spins. This number can be further reduced by a factor of 50 by increasing the resonator quality factor by matching the designed external coupling ($\mathrm{Q_e}$=3000) to the measured intrinsic quality factor $\mathrm{Q_i}$ of $\mathrm{300,000}$. Further increase in the single spin coupling from 100 Hz to 4 kHz, a factor of 40 by minimizing the air gap between the sample and the resonator, will also improve the sensitivity by a factor proportional to $\mathrm{g^2}$ = 1600. This can be done by fabricating superconducting resonators around lithographically defined RE-doped cylinders or by optimization of the sample flip-chip mounting process to minimize the air gap. Therefore, we expect to reach spin sensitivity $\mathrm{N_{spin}}$ = $\mathrm{O}$(100) by modest improvement in the device design and sample integration.

\noindent
\subsection{Angle dependence of spin inhomogeneous linewidth}
\begin{figure}
    \centering
    \includegraphics[width=0.5\textwidth]{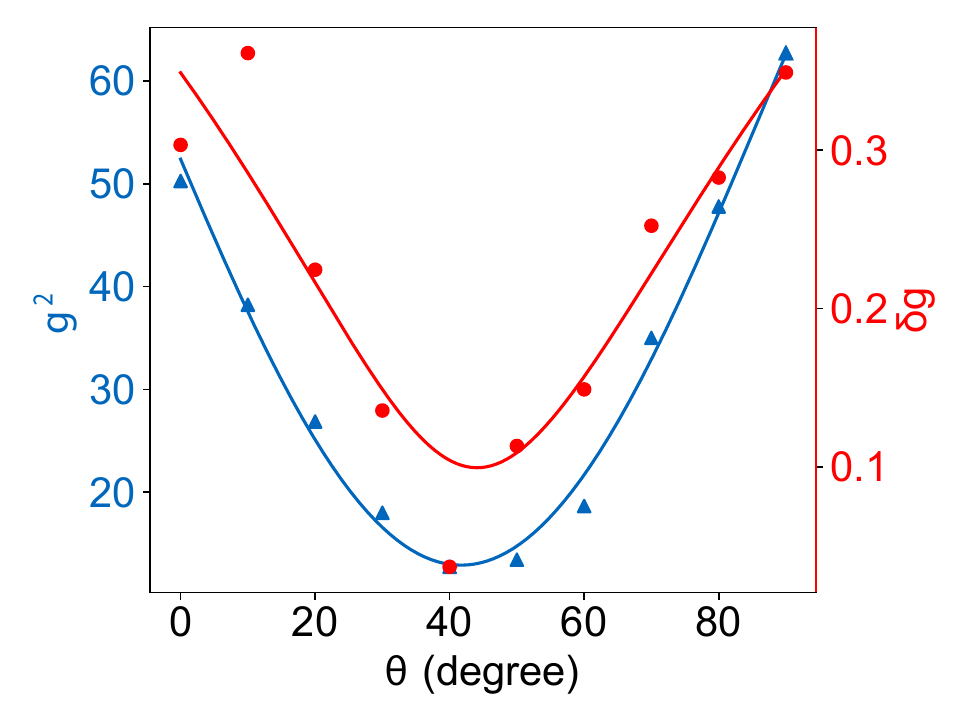}
    \caption[Angle dependence of spin inhomogeneous linewidth]{Angle dependence of spin inhomogeneous linewidth $\mathrm{\delta g}$ (red dot) and effective g factor squared $\mathrm{g^2}$ (blue triangle) for Er$\mathrm{^{3+}}$ ions in a $\mathrm{C_2}$ sub-site show close agreement with the model (blue, red line) with a similar functional dependence on the angle of the magnetic field $\mathrm{\theta}$ suggesting strain induced variation of the local environment as a mechanism of inhomogeneous broadening.}
     \label{linewidthangle}
\end{figure}

The angle dependence of the spin inhomogeneous linewidth of spins with an anisotropic g factor can be described by a perturbation of the g factor by an amount $\mathrm{\delta g}$ due to distortion of the local crystalline environment due to static strain. The square of g factor,  $\mathrm{g^2}$, and $\mathrm{g \delta g}$ follow a similar expression in this case \cite{WELINSKI201769}\cite{2022OMX....1400153L},

\begin{equation}
    \mathrm{g^2 = a \cos^2 \theta + 2b \cos \theta \sin \theta + c \sin^2 \theta }
\end{equation}
and 
\begin{equation}
    \mathrm{g \delta g = a^\prime \cos^2 \theta + 2b^\prime \cos \theta \sin \theta + c^\prime \sin^2 \theta} 
\end{equation}
where a,b,c and $\mathrm{a^\prime}$, $\mathrm{b^\prime}$, $\mathrm{c^\prime}$ are coefficients depending on the g tensor and the plane of the angle scan, $\mathrm{\theta}$ is the in-plane angle of the magnetic field. The inhomogeneous linewidths $\mathrm{\delta B}$ measured in Fig.~\ref{dispersiveabsorptive} were converted to $\mathrm{\delta g}$ following the expression $\mathrm{\delta g =  g(\delta B/B_{res}) }$, where $\mathrm{B_{res}}$ is the resonance B field. Fig.~\ref{linewidthangle} shows both the measured $\mathrm{g^2}$ and $\mathrm{\delta g}$ for a $\mathrm{C_2}$ sub-site are in close agreement with the model (red and blue lines) with both $\mathrm{g^2}$ and $\mathrm{g \delta g}$ fitted with the same functional dependence on $\mathrm{\theta}$ with different coefficients a,b,c and $\mathrm{a^\prime}$, $\mathrm{b^\prime}$, $\mathrm{c^\prime}$ respectively. A similar functional dependence of $\mathrm{\delta g}$ and $\mathrm{g^2}$ on the angle of the magnetic field $\mathrm{\theta}$ supports a strain-induced variation of the local environment as a mechanism of inhomogeneous linewidth broadening.\newline

\subsection{Estimation of magnetic noise}

The fluctuating magnetic noise amplitude ($\mathrm{\delta B}$) reported in Fig.~3 of main text was estimated using the following expression,

\begin{equation}
    \mathrm{\frac{1}{\pi T_{2,Hahn}} = S_1 \delta B + \Gamma_{res} } 
\end{equation}

\noindent
where $\mathrm{S_1}$ is the magnetic noise sensitivity of the transition given as $\mathrm{g_{eff} \mu_B}$, $\mathrm{\Gamma_{res}}$ is the residual decoherence due to instantaneous diffusion and $\mathrm{T_1}$ limit, which is estimated from the inverse of CPMG coherence time $\mathrm{\Gamma_{res}}=1/\mathrm{T_2^{CPMG}}$. $\mathrm{\delta B}$ is the RMS magnetic noise fluctuation.\newline

\subsection{Spin spectral diffusion}
To study magnetic spectral diffusion dynamics, we performed three pulses (stimulated) echo measurements for $\mathrm{C_{3i}}$ site g=3.2 transition of 130 mT at the base temperature of 12.5 mK. The waiting time ($\mathrm{T_W}$) between the second and third pulses was scanned in the range of 100 $\mathrm{\mu s}$ to 10 ms for $\mathrm{\tau}=$ 25, 30, 35, 40, 45, and 50 $\mathrm{\mu s}$. The echo amplitude decays as a function of pulse delays $\mathrm{\tau, T_W}$ can be expressed as \cite{pgoldner},

\begin{equation}
\mathrm{
    \frac{A(\tau,T_W)}{A_0)} = exp \Big[ \frac{T_W}{T_1} + 2\pi \tau \Gamma_{eff} \Big]
    }
\end{equation}

\noindent
where $\mathrm{\Gamma_{eff}}$ is the effective linewidth given as,

\begin{equation}
    \mathrm{ \Gamma_{eff} = \Gamma_0 + \frac{1}{2} \Gamma_{SD} ( R \tau + 1 -e^{-R T_W})
    }
\end{equation}

\noindent
The term $\mathrm{\Gamma_{SD}}$ represents the spectral diffusion linewidth due to dipolar coupling with environmental spin bath, R represents the effective spin-flip rate, and $\mathrm{\Gamma_{0}}$ represents instantaneous diffusion and homogeneous linewidth. In the limit of $\mathrm{R \tau} <<$ 1, we can simplify $\mathrm{ \Gamma_{eff} = \Gamma_0 + \frac{1}{2} \Gamma_{SD} (1 -e^{-R T_W})}$ \cite{erbiumSDpaper} and fit effective linewidth $\mathrm{\Gamma_{Eff}}$ for fixed $\mathrm{T_W}$ by sweeping $\mathrm{\tau}$ (inset of Fig.~4(c) in the main text). We obtained $\mathrm{\Gamma_{SD}}$ = 4.4(0.7) kHz and R= 287(140) Hz, which is comparable with slow spectral diffusion previously reported in bulk rare-earth doped crystals \cite{patrickthesis}. The 287 Hz spin-flip rate is indicative of residual flip-flops due to dipolar interaction in the paramagnetic spin bath as a source of spectral diffusion on a 10 ms timescale. However, this slow spectral diffusion does not significantly affect the linewidth on a short time scale, and the short-time-scale linewidth $\mathrm{\Gamma_{0}}$ of 0.7(0.3) kHz is close to the homogeneous linewidth of 0.8(0.03) kHz from two pulse-echo measurements. The spectral diffusion parameters $\mathrm{\Gamma_{SD}}$ and R predict a spectral diffusion limited coherence time $\mathrm{T_{2,SD} =\sqrt{\pi \Gamma_{SD}R}/2}$ of 1.01 ms. The fitted spectral diffusion parameters allow us to predict the optical linewidth broadening in the C\textsubscript{3i} site due to magnetic spectral diffusion in a 10 ms timescale as, 

\begin{equation}
        \mathrm{ \Gamma_{opt}^{mag} = \Gamma_0 + \frac{1}{2} \Gamma_{SD} ( R \tau + 1 -e^{-R (10 \, ms)}) = 2.9 \, kHz
    }
\end{equation}

\subsection{Spin decoherence mechanisms}

\begin{figure*}
    \centering
    \includegraphics[width=0.99\textwidth]{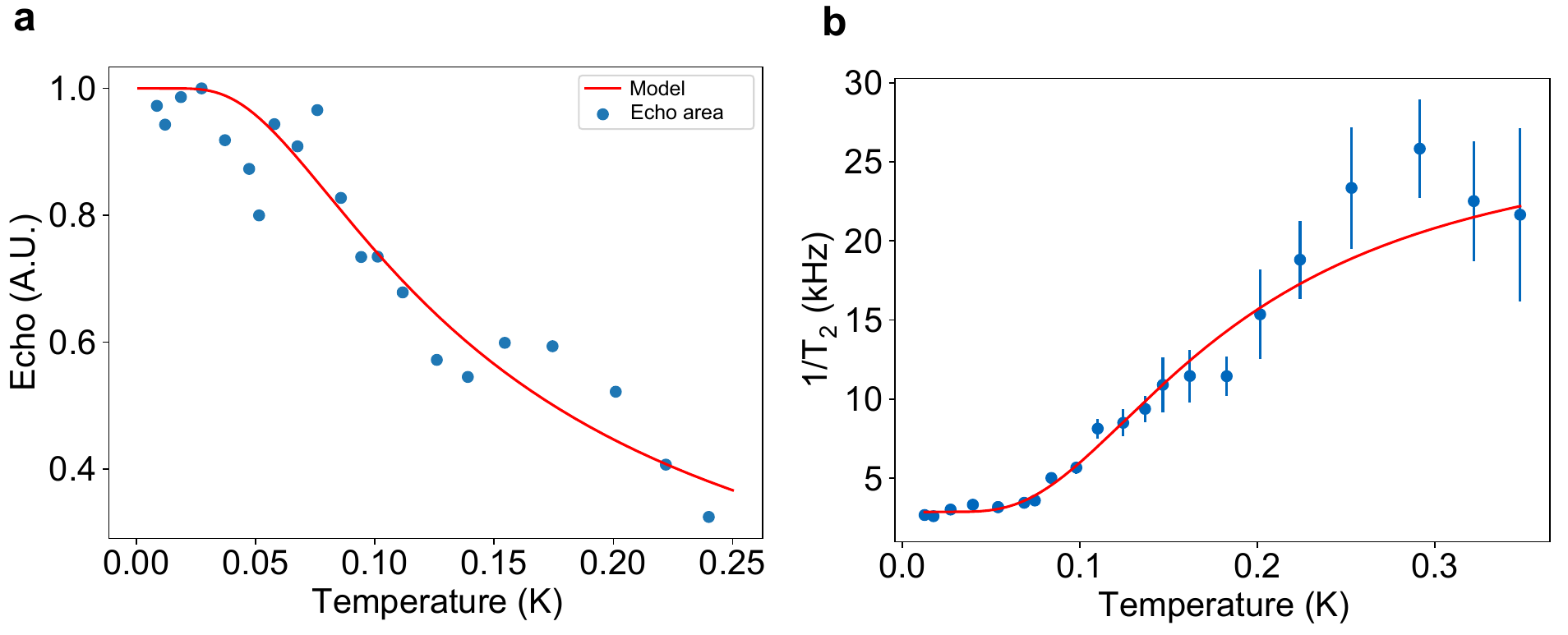}
    \caption[Temperature dependence of spin decoherence]{Spin $\mathrm{T_2}$ temperature dependence measured for the $\mathrm{C_{3i}}$ transition at 130 mT with g=3.2. {\bf a} Spin echo area corrected for $\mathrm{T_2}$ fitted to the polarization model.  {\bf b} $\mathrm{\gamma = 1/T_2}$ (blue dots) as a function of corrected temperature ($\mathrm{T_{Corr}}$) and the fitted paramagnetic spin bath induced decoherence model (red line).}
     \label{fig:tempdependent}
\end{figure*}

To study the sources of decoherence, we measure the spin coherence time ($\mathrm{T_{2}^{Hahn}}$) for the  $\mathrm{C_{3i}}$ transition at a field 130 mT, $\mathrm{\theta}$= 40 degree as a function of temperature. 
 The spin echo area for pulse separation $\mathrm{\tau}$=0 was extrapolated as $\mathrm{A(T)=A(0,T)=A(\tau,T)e^{(\tau/T_2)}}$ and fitted to temperature dependent spin polarization model (Fig.~\ref{fig:tempdependent}{\bf a}),

\begin{equation}
   \mathrm{ A(T)= A_{max} tanh (T_{Ze}/T_{Corr}) } 
\end{equation}

\noindent
where $\mathrm{A_{max}}$ is a normalization factor we set equal to 1, $\mathrm{T_{Ze}}$ is the spin Zeeman temperature set to 0.2788 K (5.8 GHz spin transition splitting), and $\mathrm{T_{Corr}}$ is the corrected spin temperature obtained by multiplying the measured sample stage temperature by a fitting parameter C, with $\mathrm{T_{Corr}= C T}$. The fit of data in Fig.~\ref{fig:tempdependent}{\bf a} to Eq.~(9) gives us a correction factor C = 1.45(0.06). A good agreement between the data and model validates the fitting and provides a corrected base temperature of 12.5 mK. At this temperature, we expect a unity spin polarization (p) at 5.8 GHz: $\mathrm{p = tanh(hf/2k_BT)}$ and 1-p $\approx \mathrm{ 9 \times 10^{-20}}$. 

After correcting the spin temperature, the temperature dependence of the dephasing rate $\mathrm{1/T_2}$ was modeled as \cite{multimodeeryso} \cite{PhysRevLett.101.047601},
\begin{equation}
    \mathrm{\frac{1}{T_2(T)} = \Gamma_{0} + \frac{\xi}{(1+ e^{T_{Ze}/x})(1+ e^{-T_{Ze}/x})}}
\end{equation}

\noindent
where $\mathrm{\Gamma_{0}}$ is a temperature-independent part of decoherence such as instantaneous diffusion, $\mathrm{\xi}$ is a temperature-independent parameter with a strength proportional to the dipolar coupling to paramagnetic spin bath, and $\mathrm{T_{Ze}}$ is the Zeeman temperature of the spin bath. We fit the model to the data in Fig.~\ref{fig:tempdependent}{\bf b} to obtain $\mathrm{\xi}$= 96.8(9.7) kHz, $\mathrm{\Gamma_0}$=  2.9(0.1) kHz and $\mathrm{T_{Ze}}$ = 0.34 K. The Zeeman temperature $\mathrm{T_{Ze}}$ = 0.34 K corresponds to a spin bath splitting of 6.9 GHz (equivalently g=3.8) and supports $\mathrm{Er^{3+}}$ in $\mathrm{C_2}$ sites with g $\approx$ 3.8 as the primary source of decoherence at higher temperatures (T$\geq$ 0.1 K). Indeed, out of the six $\mathrm{C_2}$ sub-sites, two orientations have a g= 3.6, and four orientations have g= 8.6 for the in-plane magnetic field direction $\mathrm{\theta}$=40 degrees in this measurement. 
We also note that the theoretical dipolar coupling of the Er\textsuperscript{3+} spins in $\mathrm{C_{3i}}$ site to the $\mathrm{C_{2}}$ site spin-bath with a total concentration of $\mathrm{n_{C_{2}}}$ of 81.4 ppm and g=3.8 is $\mathrm{\nu_{dd} =  (h \mu_0 \gamma_{C_{3i}}\gamma_{C_{2}} n_{C_{2}})/(4\pi)}$ = 295 kHz, which is close to experimentally estimated parameter $\xi$ = 96.8 kHz within a factor of three \cite{patrickthesis}. The discrepancy of a factor of 3 can be accounted for given that $\xi$ does not have a closed-form expression and anisotropy of g factor leads to a complex dipolar interaction \cite{pgoldner} \cite{milliKelvinESR}.

\begin{figure}
    \centering
    \includegraphics[width=0.5\textwidth]{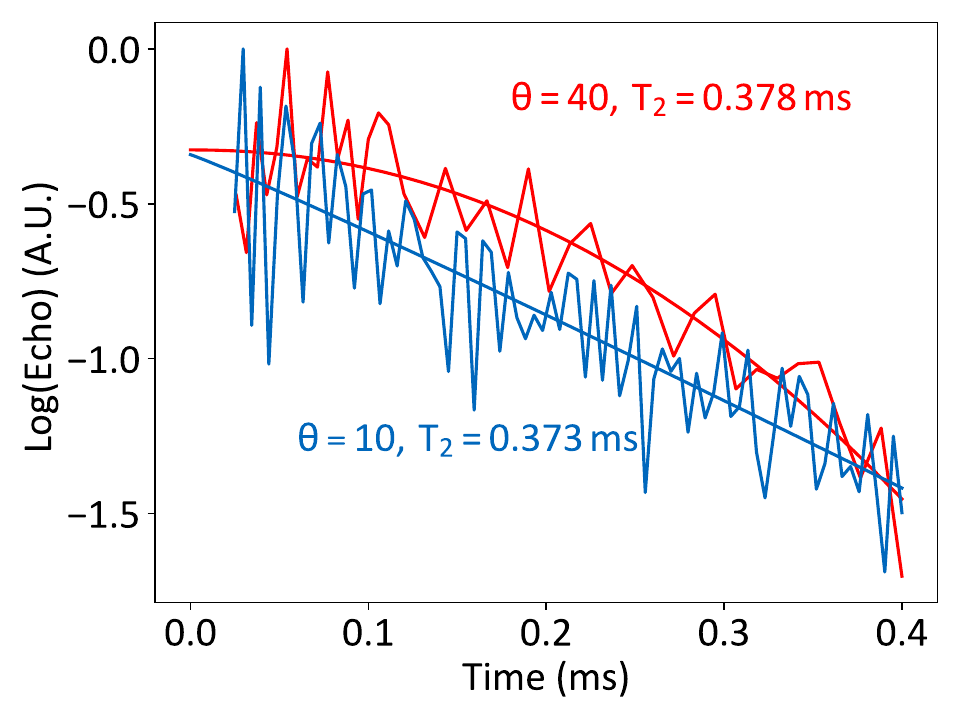}
    \caption[Angle dependence of spin coherence time]{ Similar spin coherence time ($\mathrm{T_2^{Hahn}}$)  was measured for the $\mathrm{C_{3i}}$ g=3.2 transition for different in-plane angle $\mathrm{\theta}$ corresponding to different g factor of $\mathrm{C_2}$ sub-sites, thereby confirming the suppression of flip-flops of $\mathrm{Er^{3+}}$  $\mathrm{C_2}$ site spin-bath at the base temperature of 12.5 mK.}
    \label{fig:t2vstheta}
\end{figure}

The saturation of the dephasing rate $\mathrm{1/T_2}$ for temperatures less than 0.1 K indicates suppression of Er\textsuperscript{3+} C\textsubscript{2} site spin-bath flip-flops due to near unity spin polarization. To verify this, we repeat the $\mathrm{T_2}$ measurement on the same $\mathrm{C_{3i}}$ transition with g=3.2 at a different in-plane angle $\mathrm{\theta}$ = 10 degrees where all the $\mathrm{C_2}$ subsites have a different g = 6.2 (all six orientations) or equivalently a splitting of 11 GHz at 132 mT. Observation of similar coherence times for different $\mathrm{C_2}$ site spin-bath splittings would rule out $\mathrm{C_2}$ site as the source of decoherence. We observe near identical spin coherence times ($\mathrm{T_2^{Hahn}}$) of 0.373 (0.038) ms and 0.378 (0.012) ms for the $\mathrm{C_{3i}}$ transition corresponding to $\mathrm{\theta}$ = 10 and $\mathrm{\theta}$ = 40 degrees respectively at 12.5 mK temperature (Fig.~\ref{fig:t2vstheta}). Changing the $\mathrm{C_2}$ site spin-bath splitting by applying field at different angles does not affect the spin coherence time. This confirms that the $\mathrm{Er^{3+}}$ $\mathrm{C_2}$ site spin-bath is polarized at the base temperature of 12.5 mK, and the low-temperature dephasing rate $\mathrm{\Gamma_{0}}$= 2.9 kHz is dominated by temperature-independent processes (such as spin-spin interaction) and residual magnetic noise. 

We estimate 1 kHz contribution of spin-spin interaction (instantaneous diffusion) to $\mathrm{\Gamma_{0}}$ based on CPMG measurements ($\mathrm{\Gamma_{ID}}$=$\mathrm{1/T_2^{CPMG}}$), which allows us to put a limit on decoherence due to residual magnetic noise as $\mathrm{\Gamma_{Res, Mag} = \Gamma_{0}- \Gamma_{ID}}$ = $\mathrm{1/(526 \, \mu s)^{-1}}$. The residual magnetic noise at the base temperature could be due to a combination of superhyperfine contribution from yttrium nuclear spins \cite{y2o3millisecond2022} and traces of unpolarized paramagnetic impurities and defects (Section \ref{TLSsection} and Fig.~\ref{TLS}) \cite{y2o3millisecond2022}. Thus, with a decrease in Er\textsuperscript{3+} and paramagnetic defect/impurity densities, we expect further reduction in magnetic noise, spin-spin interaction, and hence improvement in spin coherence times at sub-Kelvin temperatures.

\subsection{Instantaneous diffusion and inhomogeneous linewidth at the wafer-scale}
\label{inhomcpmgsection}

\begin{figure}
    \centering
    \includegraphics[width=0.99\textwidth]{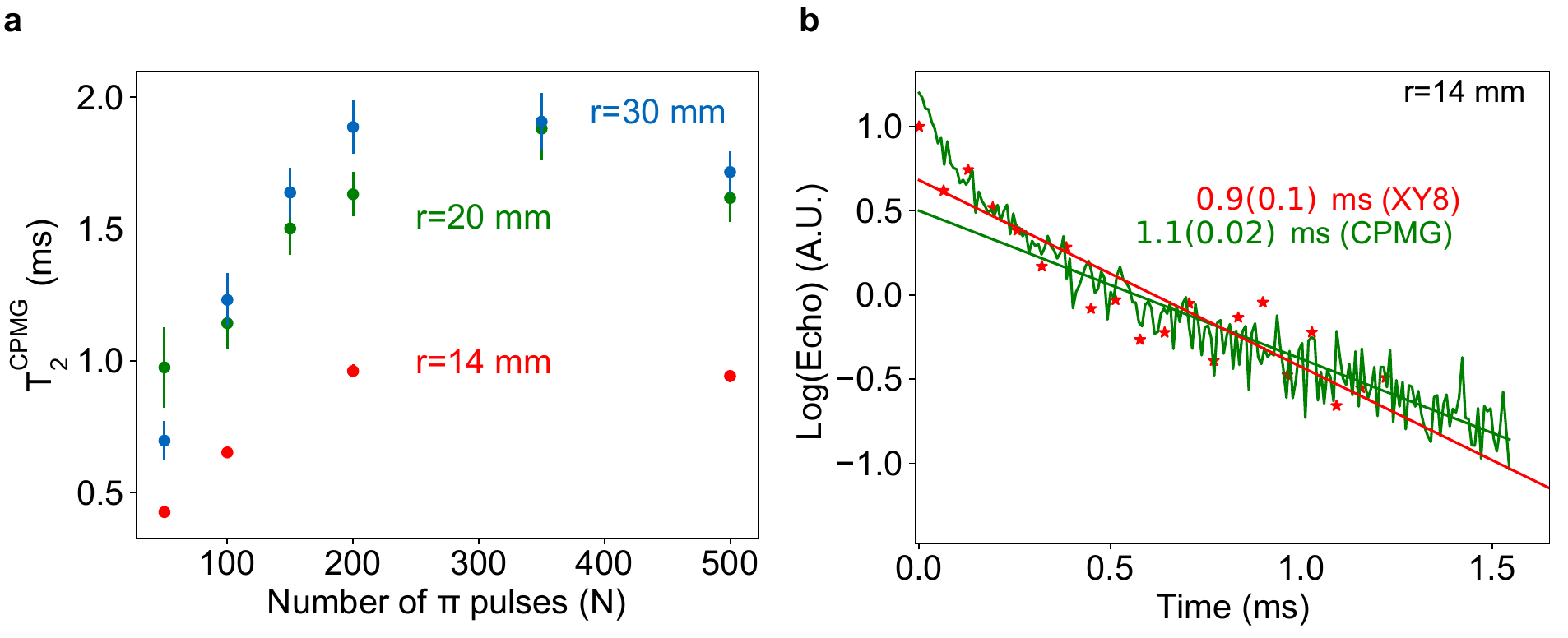}
    \caption[Spin coherence time under dynamical decoupling]{Spin coherence time under dynamical decoupling. {\bf a} Spin coherence time with CPMG dynamical decoupling pulses ($\mathrm{T_2^{CPMG}}$) as a function of the number of pulses (N) for samples taken from different positions (r) on the wafer. A higher saturation value of $\mathrm{T_2^{CPMG}}$ for larger r suggests a reduction in spin spectral density with r. {\bf b} Similar spin coherence times were observed under the CPMG (green) and XY8 (red) sequence with N=200 $\pi$ pulses indicating an instantaneous diffusion limit.}
    \label{fig:cpmgvsN}
\end{figure}

\noindent
\textbf{Instantaneous diffusion limit under dynamical decoupling} To study the instantaneous diffusion limit, we performed CPMG dynamical decoupling measurements as a function of the number of pulses N for samples taken from different radial positions (r) along the wafer (Fig.~\ref{fig:cpmgvsN}{\bf a}) at 12.5 mK. While all samples follow the trend of an initial linear increase in $\mathrm{T_2^{CPMG}}$ with the number of $\pi$ pulses followed by a saturation after N=200, the samples towards the edge of the wafer (large r) show a greater $\mathrm{T_2^{CPMG}}$. We also note a slight decrease in $\mathrm{T_2^{CPMG}}$ for N= 400 and 500 pulses, which could be attributed to the accumulation of pulse errors \cite{dynamicdecouplingcolloq}. Increasing the number of pulses can lead to efficient decoupling from the spin bath due to a narrower filter function. However, the saturation of $\mathrm{T_2^{CPMG}}$ with N suggests instantaneous diffusion between resonant spins as the limiting source of decoherence \cite{quantummetrology}. Further, repeating dynamical decoupling measurement with a more robust XY8 sequence (Fig.~\ref{fig:cpmgvsN}{\bf b}, red) produces similar coherence times as CPMG sequences (Fig.~\ref{fig:cpmgvsN}{\bf b}, green), thus further evidencing instantaneous diffusion as the limiting decoherence mechanism. This allows us to estimate the effective spin concentration $\mathrm{n_{eff}}$ from the ID-limited CPMG coherence time as,

\begin{equation}
    \mathrm{n_{eff}= \frac{9 \sqrt{3}}{2 \pi^2 \mu_0 h \gamma^2 T_2^{CPMG}}}
    \label{eqnneff}
\end{equation}

\noindent
The higher saturation value of $\mathrm{T_2^{CPMG}}$ for large r implies a decrease in instantaneous diffusion for the samples closer to the edge. This could be explained by an observed increase in spin inhomogeneous linewidth towards the edge, leading to a decrease in spin spectral density, in close agreement with the increase in T\textsubscript{1} towards the wafer edge. \newline

 \begin{figure}
    \centering
    \includegraphics[width=0.5\textwidth]{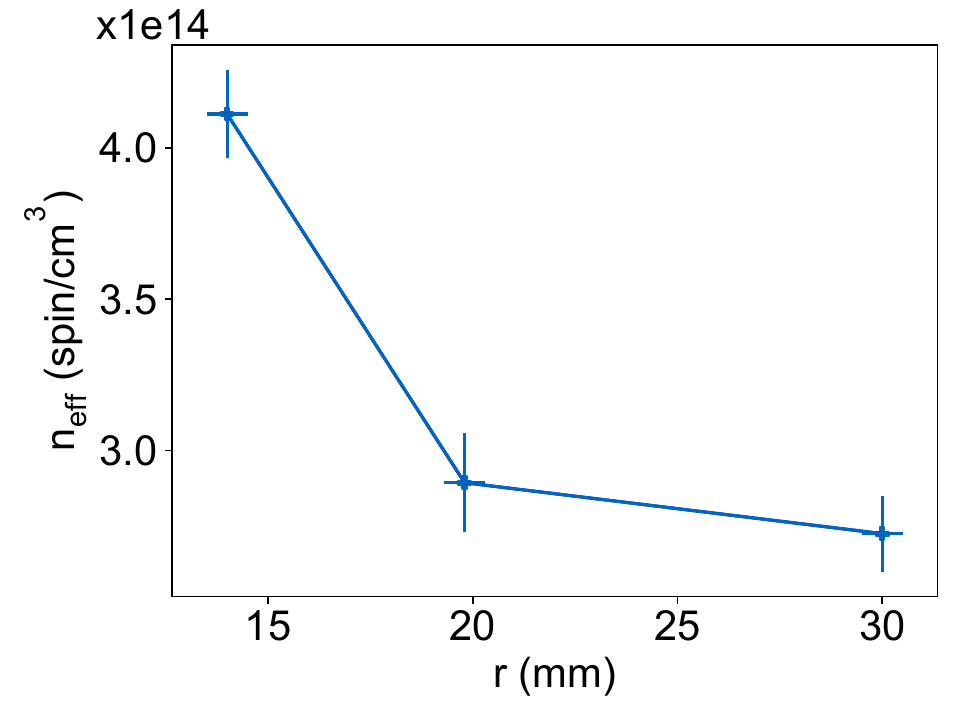}
    \caption[Effective spin density on a wafer-scale]{Estimated effective $\mathrm{Er^{3+}}$ spin density ($\mathrm{n_{eff}}$) versus position on the sample (r) deduced from the CPMG data using Eq.~\ref{eqnneff}. A reduction in effective density with r is in agreement with the increase in inhomogeneous linewidth in Fig.~\ref{fig:gammavsr}.}
    \label{fig:nvsr}
\end{figure}

\begin{figure*}
    \centering
    \includegraphics[width=1.0\textwidth]{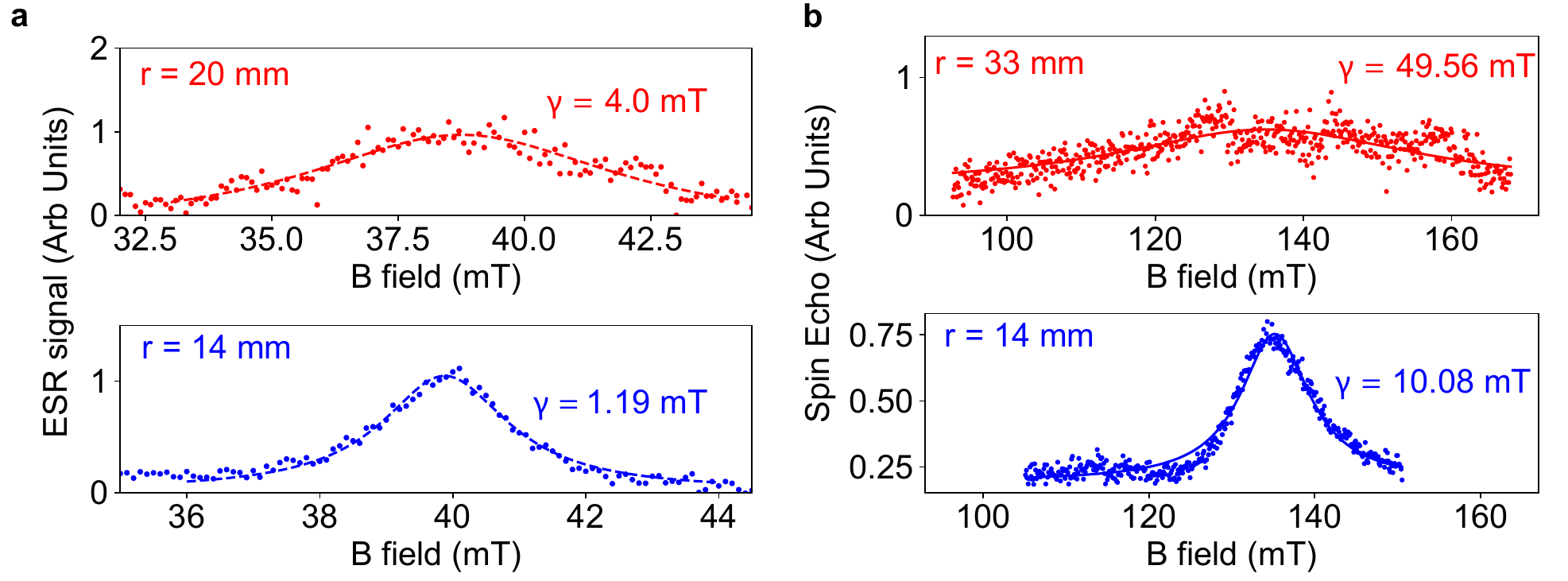}
    \caption[Inhomogeneous linewidth on wafer scale]{Comparison of spin inhomogeneous linewidths as a function of distance (r) from the center of the wafer for low and high field transitions. {\bf a} The inhomogeneous linewidth of a low field transition for the sample closer to the center (r=14 mm, blue) compared to the sample farther from the center (r= 20 mm, red). The $\mathrm{Er^{3+}}$ spin ensemble far from the center (red) shows at least 3.43 times larger inhomogeneous linewidth than the broadest linewidth measured for a similar transition for the sample close to the center (blue). {\bf b} The inhomogeneous linewidth for a high field transition for another sample even farther from the center (r= 33 mm, red) compared with the same sample closest to the center (r=14 mm, blue). This sample shows a 5 times broader inhomogeneous linewidth than the sample near the center (blue).}
    \label{fig:gammavsr}
\end{figure*}

\noindent
\textbf{Inhomogeneous linewidth on wafer-scale} Figure~\ref{fig:gammavsr} compares the spin inhomogeneous linewidth for both low and high field transitions between samples taken from different points of the wafer (r). The low field transition (Fig.~\ref{fig:gammavsr}{\bf a}) measured using CW ESR shows 3.4 times broader linewidth for the sample farther from the center (r=20 mm, red) than the sample close to the center (r=14 mm, blue). The high field transition measured using spin echo detection (Fig.~\ref{fig:gammavsr}{\bf b}) also shows a broader inhomogeneous linewidth by a factor of 5 for the sample at r=33 mm than the sample at r=14 mm. Therefore, a significant increase in spin inhomogeneous linewidth with distance (r) for both low and high field transitions explains a reduced spin spectral density and hence reduced instantaneous diffusion, which is in close agreement with measurements in Fig.~\ref{fig:cpmgvsN}. This increase in inhomogeneous linewidth towards the edge, along with the modeled strain-induced angle dependence of linewidth in Fig.~\ref{linewidthangle}, indicates an increase in lattice strain towards the edge.\newline

\noindent
 \textbf{Theoretical instantaneous diffusion limit} To estimate the theoretical ID limit, we first estimate the number of photons corresponding to -65 dBm on-chip power as $\mathrm{\langle n \rangle = (P/hf) \times (ke/k^2) }$ = 41085 using f = $\mathrm{5.8 \times 10^9 }$ Hz, $\mathrm{k \approx k_e = }$ 1.9 MHz for the device used in Fig.~2 in the main text. We then estimate the Rabi frequency as $\mathrm{\Omega= g \sqrt{\langle n \rangle}}$ $\sim 0.78 $ MHz using $g \sim 128 $ Hz. 
 The ID limit can, therefore, be estimated as,

 \begin{equation}
    \mathrm{T_{2,ID}^{-1}= \frac{2\pi^2 \mu_0 h n \gamma^2}{9 \sqrt{3}}  \langle sin^2 \frac{\theta}{2} \rangle } 
\end{equation}

\noindent
where $\mathrm{\langle sin^2 \frac{\theta}{2} \rangle}$ is the average spin-flip probability evaluated by integrating the spin-flip probability over the inhomogeneous linewidth \cite{pgoldner}. Using inhomogeneous linewidth of 100 MHz, Rabi frequency $\mathrm{\sim}$ 0.78 MHz, we get $\mathrm{\langle sin^2 \frac{\theta}{2} \rangle} \sim$ 0.016. Using these values and the estimated $\mathrm{C_{3i}}$ density of $\mathrm{32.4 \times 10^{16} spins/cm^{3}}$ we get $\mathrm{T_{2,ID}} \sim
$ 0.50 ms. The measured $\mathrm{T_{2}^{CPMG}}$ $\approx$ 1.1 ms is consistent with the theoretical lower bound on ID, and the largest source of uncertainty in this calculation is the estimated single spin coupling g.

\begin{center}
\begin{table}
\begin{tabular}{||c c c c c||} 
 \hline
 Device & $\mathrm{Q_{i0}}$ & $\mathrm{F tan \delta}$ & $\mathrm{nc}$ & $\mathrm{\alpha}$\\ [0.5ex] 
 \hline\hline

 1(r=20 mm) & 13422 & $\mathrm{3.92 \times 10^{-5}}$ & $\mathrm{3.8 \times 10^{6}}$ & 0.36 \\ 
 \hline

 2(r=30 mm) & 14088  & $\mathrm{3.10 \times 10^{-5}}$ & $\mathrm{2.6 \times 10^{5}}$ & 0.54 \\
 \hline
 
  3(r=33 mm) & 13082 & $\mathrm{3.78 \times 10^{-5}}$ & $\mathrm{1.1 \times 10^{6}}$ & 0.54\\[1ex] 
 \hline
 \hline
\end{tabular}
\caption{Power dependence fit parameters for superconducting devices used in Fig.~3 of the main text.}
\label{tlstable}
\end{table}
\end{center}

\subsection{ Spin T\textsubscript{1} dependence on temperature and resonator frequency detuning}
\label{sectspinT1}

\begin{figure*}
    \centering
    \includegraphics[width=0.99\textwidth]{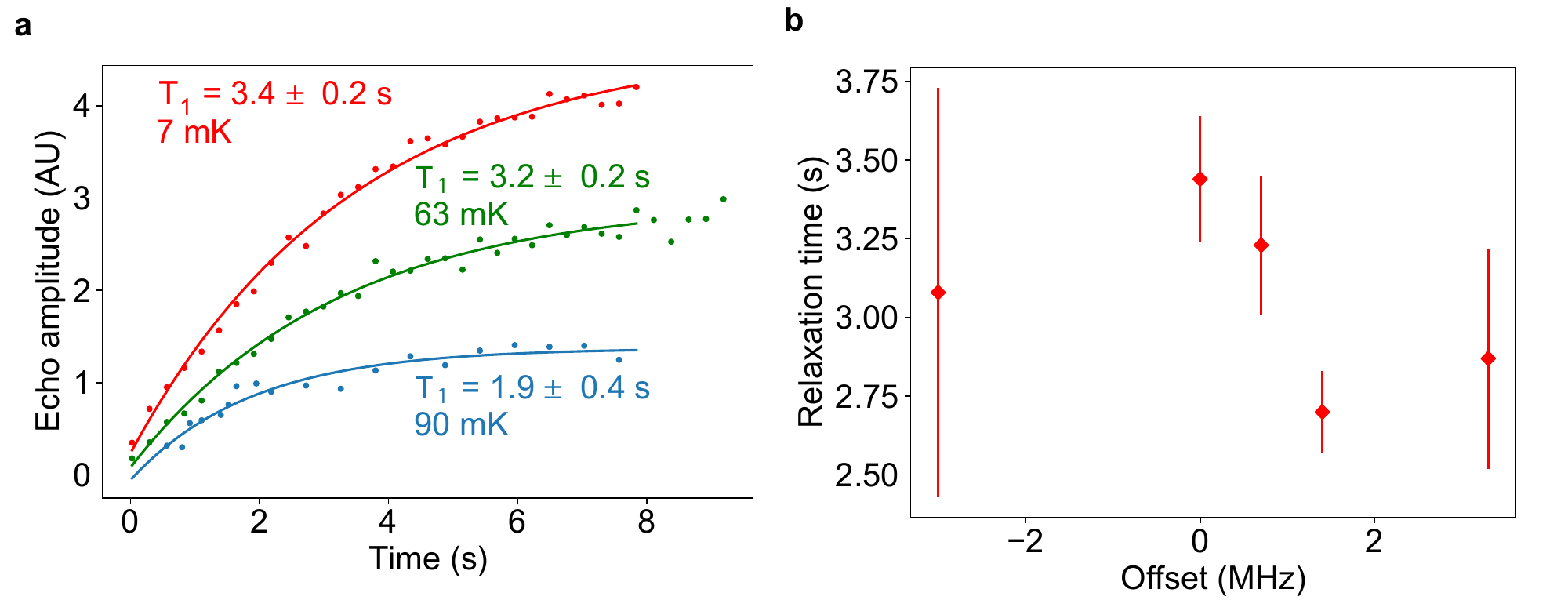}
    \caption[Spin T\textsubscript{1} versus temperature and offset.]{Spin T\textsubscript{1} versus temperature and microwave-cavity frequency offset. {\bf a} Spin T\textsubscript{1} of the C\textsubscript{2} g=3.6 transition measured as a function of the recorded temperature. {\bf b} Spin T\textsubscript{1} at the base temperature as a function of offset between microwave source and resonator center frequency. Resonator linewidth $\kappa$ was 1.94 MHz.}
     \label{t1vsofftemp}
\end{figure*}

Spin T\textsubscript{1} measured for the g=3.6 transition from the C\textsubscript{2} sub-site shows a weak dependence on temperature (Fig.~\ref{t1vsofftemp} {\bf a}), with a small decrease from 3.4 s at 7 mK to 1.9 s at 90 mK. Weak temperature dependence of spin T\textsubscript{1} is expected for direct process-dominated spin relaxation where the relaxation rate  ($\mathrm{\Gamma_{D} \propto \coth{(hf/2k_BT)} }$) saturates at sufficiently low temperatures ($\mathrm{T << hf/2k_B \approx} $ 286 mK). 

 Spin T\textsubscript{1} at 7 mK further shows no dependence on the microwave frequency detuning from the resonator center frequency (Fig.~\ref{t1vsofftemp}{\bf b}), even for detunings larger than the resonator linewidth $\mathrm{\kappa = 1.94 }$ MHz. This allows us to constrain the Purcell-enhanced spin relaxation rate $\mathrm{\Gamma_P \leq 0.5*(2.75s^{-1} -3.4 s^{-1}) = 0.034 \, Hz }$. Using the expression $\mathrm{\Gamma_P = 4 \kappa g^2/(\kappa^2 + \delta^2)}$ \cite{JpettaPhysRevLett.118.037701}, and substituting for $\mathrm{\Gamma_P \leq 0.034 \, Hz}$, $\mathrm{\kappa = 1.94 \, MHz, \delta = 0}$, we get spin-resonator coupling strength as g \textless 128 Hz. While Purcell enhanced Er\textsuperscript{3+} spin relaxation has been previously reported in ESR studies with superconducting resonators \cite{doi:10.1126/sciadv.abj9786}, the lack of such observation in our measurements can be explained by two factors. Firstly, the presence of large air gaps $\gtrapprox$ 1-2 $\mu m$ between the $\mathrm{Y_2 O_3}$ thin-film and superconducting resonator leads to a reduction of spin-resonator coupling strength g, as shown in Fig.~\ref{fig:resonatorsimulationl}. These air gaps can be removed by optimization of the flip-chip mounting scheme. Further, superconducting resonators can be patterned around lithographically defined yttrium oxide thin-film cylinders. With these improvements, single spin coupling g as high as 4 kHz is expected. Secondly, a relatively high Er\textsuperscript{3+} density of $\approx$ 12-13 ppm in our measurement as compared to $\approx$ 1 ppb density in ref.~\cite{doi:10.1126/sciadv.abj9786} leads to significant spin-spin relaxation which competes with Purcell enhanced spontaneous emission. 

\subsection{ Increasing Spin T\textsubscript{1} with lower doping density and magnetic fields} 
\noindent Direct-process limited relaxation rate scales as $\mathrm{B^{-5}}$ at low temperatures while cross-relaxation limited spin relaxation scales as the square of density ($\mathrm{n^2}$). Therefore, long spin lifetimes can be expected by operating at lower magnetic fields and/or lower $\mathrm{Er^{3+}}$ density\cite{PhysRevB.95.205119}. Indeed, $\mathrm{T_1}$ of 15 s limited by the direct process was observed for a low density(0.7 ppb) of $\mathrm{Er^{3+}}$ in $\mathrm{CaWO_4}$ (10 mK, 500 mT)\cite{doi:10.1126/sciadv.abj9786}. Low magnetic field measurements of $\mathrm{Er^{3+}}$ in $\mathrm{Y_2SiO_5}$ have demonstrated direct process limited $\mathrm{T_1}$ of 45 s (0.5 K, 11 mT) \cite{Raha2020} and even longer phonon-bottlenecked direct process limited $\mathrm{T_1}$ of 10 hours at lower fields and temperature (20 mK, 6.5 mT) \cite{Budoyo_2018}. This indicates a possibility of significantly increasing the spin lifetime by operating at lower $\mathrm{Er^{3+}}$ densities and magnetic fields, typically used in experiments on optically addressed single RE qubits \cite{Kindem2020} \cite{Raha2020}.

\subsection{Magnetic and electrical TLS in rare-earth/superconducting hybrid device}
\label{TLSsection}

\begin{figure*}[t!]
    \centering
         \includegraphics[width=1\textwidth]{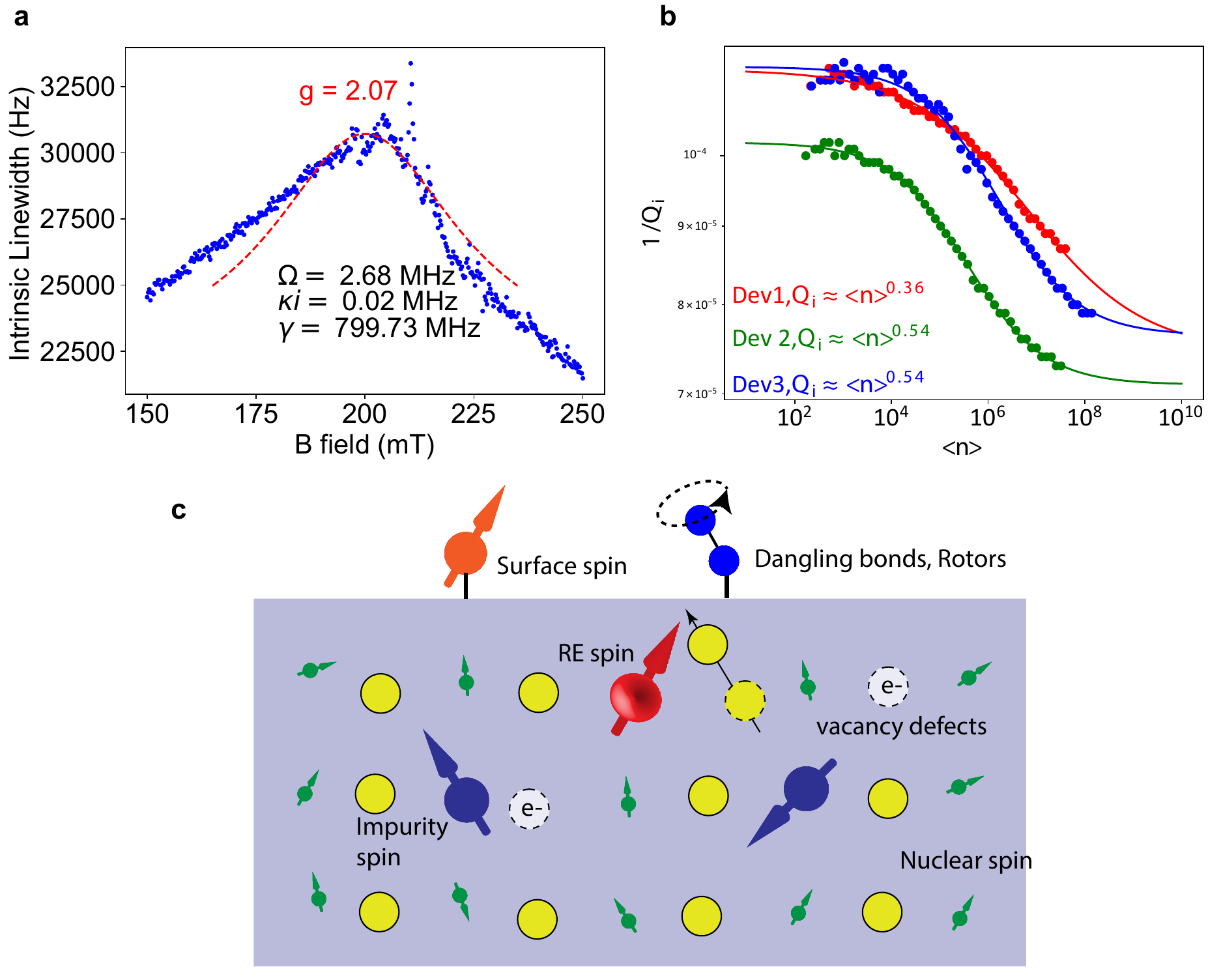}
    \caption[Magnetic and electrical TLS in erbium-resonator hybrid device]{Magnetic and electrical TLS in rare-earth/superconducting hybrid device. {\bf a} A broad g=2 feature was observed along with a smaller satellite peak CW-ESR scans, suggesting the presence of magnetic TLS for the device used in Fig 2. The magnetic TLS could be attributed to paramagnetic defects or impurities in the bulk of material or surface spins. {\bf b} Electrical TLS: Power-dependent quality factor measurements on three different devices from Fig. 3 indicate the presence of electrical TLS. {\bf c} Noise sources in rare-earth/superconducting hybrid devices: Magnetic TLS, such as paramagnetic impurities, surface spins, defects, Er in different sub-sites, and electrical TLS, such as dangling bonds and vacancy defects, can be a significant source of loss and decoherence in hybrid devices.} 
    \label{TLS}
\end{figure*}

\noindent
{\bf Magnetic TLS} Figure~\ref{TLS}{\bf a} shows a broad g $\mathrm{\approx}$ 2.07 background signal observed in the form of an increase in resonator linewidth when performing CW ESR on the device used in Fig.~2 in the main text. The signal had a broad asymmetric lineshape, which could be explained by the presence of a broad background of multiple paramagnetic sub-species with g $\mathrm{\approx}$ 2. Fitting the model $\kappa_i = \kappa_{i,0} + \Omega^2 \gamma_s/ (\gamma_s^2+ \Delta^2)$ provides an ensemble coupling $\mathrm{\Omega}$ of 2.68 MHz and a spin half-width of $\mathcal{O}$(800 MHz). This allows us to estimate a total number of 15 $\mathrm{\times 10^9}$ paramagnetic spins distributed over the inductor area of $\mathrm{\approx}$ 1256 $\mathrm{\mu m^2}$. Therefore, the magnetic TLS has an area density of 12.5 $\times \mathrm{10^6}$ spins/$\mathrm{\mu m^2}$. Assuming the TLS are localized in the thin film with 1.5 $\mathrm{\mu m}$ thickness, we get a total volume density of magnetic TLS as $\mathrm{8.32 \times 10^{18} \, spin/cm^3}$. 
These defects could be paramagnetic impurities or defects like charged oxygen vacancies ($\mathrm{F^+}$ center) and interstitial defects (O\textsuperscript{2-}) which have been previously reported in $\mathrm{Y_2 O_3}$ thin-films as well as ceramics \cite{Kunkel2016} \cite{y2o3millisecond2022}. Another possible contribution to the magnetic TLS signal could be from silicon substrate or niobium surface oxides. Therefore, the estimated density is the upper limit on magnetic defect density in the thin film. 


Angle dependence of inhomogeneous linewidth (Fig.~\ref{linewidthangle}) indicates strain as a dominant mechanism of inhomogeneous linewidth broadening. Relaxation of strain has been shown to be accompanied by the creation of point defects, such as oxygen vacancies in epitaxial thin films, which could contribute to decoherence \cite{PhysRevB.88.054111} \cite{Herklotz_2017}. An increase in $\mathrm{Er^{3+}}$ inhomogeneous linewidth towards the edges of the wafer as measured in Fig.~\ref{fig:gammavsr} indicates an increase in strain. While magnetic TLS measurement was not performed on the samples farther from the center, we do not see a change in $\mathrm{Er^{3+}}$ spin $\mathrm{T_2}$ or homogeneous linewidth across different samples, which indicates a small change in paramagnetic defects density towards the edges. Thus, strain does not play a strong role in increasing the density of paramagnetic defects in this measurement.\newline

\noindent
{\bf Electrical TLS}  Power-dependent quality factor measurements of superconducting resonators can be used to study the density and dynamics of electrical TLS. We performed power dependence measurements on the three resonators used in Fig.~3 of the main text. This allows us to obtain the variation of TLS density and interaction strength throughout the thin-film sample. Fig.~\ref{TLS}{\bf b} shows the inverse of intrinsic quality factor ($\mathrm{Q_i}$) as a function of the number of photons in the microwave resonator $\mathrm{\langle n \rangle}$. The data were fitted to the model \cite{deGraaf2018},
  \begin{equation}
      \mathrm{ \frac{1}{Q_i} = \frac{1}{Q_{i0}} + \frac{F tan \delta}{(1+(x/nc)^\alpha)}}
  \end{equation}

\noindent
where $\mathrm{Q_{i0}}$ is the quality factor due to power-independent losses. In the low power limit, $\mathrm{F tan \delta}$ is proportional to the TLS loss tangent $\mathrm{\delta}$ times the filling factor F in the resonator, $\mathrm{nc}$ is the critical photon number for saturation and $\mathrm{\alpha}$ is the exponent of the power dependence. The fitting parameters for all 3 devices are shown in Table \ref{tlstable}, and the data with fit in Fig.~\ref{TLS}{\bf b}. All devices exhibited a similar power-independent internal quality factor ($\mathrm{Q_{i0}}$) between 13,000-14000 as well as similar $\mathrm{F tan \delta}$, suggesting a similar density of electrical TLS on RE/superconducting surfaces across the wafer scale. A decrease in $\mathrm{\alpha}$ to 0.36 for Dev 1 is indicative of interacting TLS leading to weaker power dependence of the quality factor for that device. 
 
We also observe a signature of coherent electrical TLS as a background echo signal for all devices when measuring at a frequency off-resonant to the superconducting resonator. This echo is attributed to coherent TLS coupling to the microwave transmission line. These TLS echoes exhibit no magnetic field dependence in the 0-0.5 Tesla range, and we obtained a TLS $\mathrm{T_2} \approx$ 1.5 $\mathrm{\mu s}$ and a short TLS $\mathrm{T_1}$ of 5.6 $\mathrm{\mu s}$. The TLS signal was also observed in a range of devices, including the devices with Nb patterned directly on $\mathrm{Y_2 O_3}$ as well as the flip-chip mounted devices. The latter devices had an insignificant overlap of the rare-earth doped thin film with the microwave waveguide field, which suggests the electrical TLS could originate from niobium surface oxides. While the TLS show coherent properties, their presence poses a challenge for detecting the spin echo signal over a non-zero background echo signal, and it could also play a role in inducing resonator loss as well as decoherence of rare-earth dopants due to the electrical noise created by fast TLS $\mathrm{T_1}$ relaxations.

Overall, the presence of magnetic and electrical TLS indicates further improvement in device performance and spin coherence time is feasible, for instance, by optimizing the annealing conditions to heal bulk defects in $\mathrm{Y_2 O_3}$ and the passivization techniques to remove surface spins on rare earth/niobium interfaces.

\section{Optical coherence spectroscopy}
\renewcommand{\thesubsection}{2.\arabic{subsection}}

\subsection{Optical measurement setup}
\begin{figure*}[t!]
    \centering
    \includegraphics[width=1.0\textwidth]{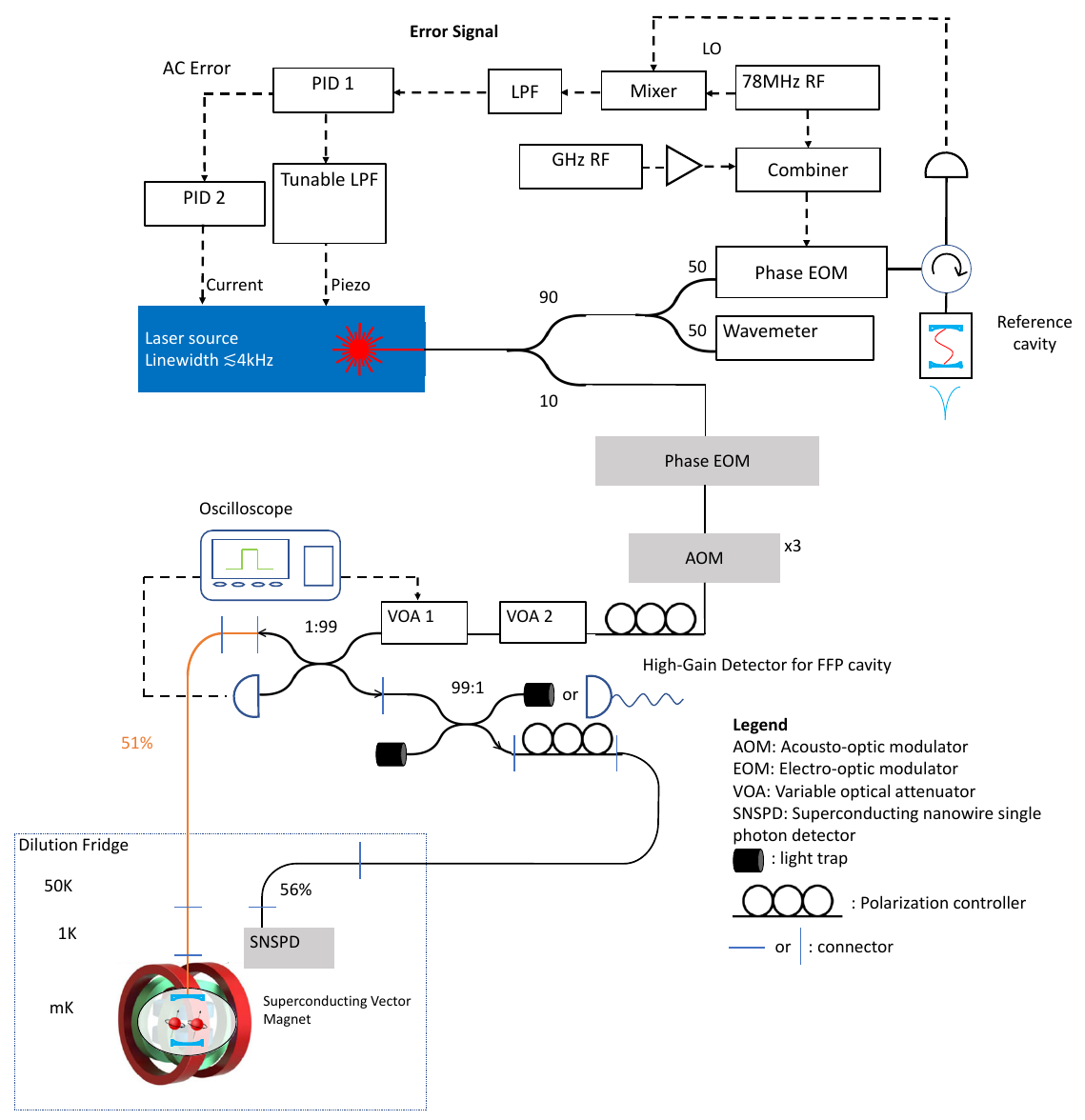}
    \caption[Optical spectroscopy setup]{ Optical spectroscopy setup. The laser frequency is locked to a stable reference cavity, and the power is actively stabilized using two variable optical attenuators. The laser is modulated using AOMs for pulse shaping, and the photoluminescence signal is detected using an SNSPD.}
    \label{fig:opticalsetpu}
\end{figure*}

\noindent As shown in Fig.~\ref{fig:opticalsetpu}, the optical power of $\approx$30 mW provided by a continuously tunable diode laser (Toptica CTL1500) was split into three paths, one for laser locking, one for wavelength monitoring, and the remaining (10 \%  power) was sent to the optical spectroscopy setup.

The laser locking setup utilized the Pound-Drever-Hall (PDH) technique to lock the laser to a UHV stable reference cavity. A phase electro-optic modulator (EOM) was placed in the locking path for laser modulation. The RF input of the phase EOM consisted of a LO signal at $\omega_{SB} \approx$ 100-800 MHz (Windfreak Synth NV), which was further combined with $\omega_{dith}$ $\approx$ 78 MHz dithering signal (Siglent SDG 2122 X) using a radio frequency combiner. Upon modulation, the EOM produced two laser sidebands around the laser carrier at frequency $\omega_0 \pm \omega_{SB}$, with each sideband modulated at the dithering frequency $\omega_{dith}$. The EOM was driven at high voltages to suppress the carrier peak in comparison to the first-order sidebands. The modulated laser was passed to a circulator on port 1, which was further routed to a reference cavity on port 2 (SLS  VH-6010-4), which has a linewidth of 150 kHz and free spectral range (FSR) of 1.5 GHz. The signal reflected from the cavity was measured on a photodetector (Thorlabs APD430C). The output of the photodetector is further mixed (Minicircuits ZP-3MH+) with the reference dither signal at frequency $\omega_{dith}$ to produce the error signal. Thus, upon mixing, we expect dispersive error signals at the frequencies $\omega_0$ and $\omega_0  \pm \omega_{SB}$. The error signal was split into two paths, to a PID controller (PID1, Vescent Photonics D2-125) where the DC part of the error signal was filtered by a low pass filter (Stanford Research Systems Model SR640) to provide a piezo feedback to the laser. The AC error signal was fed into a second PID controller (PID2, New focus LB1005), which provided current feedback to the laser. The laser was locked to the error signal from either of the two sidebands $\omega_0 \pm \omega_{SB}$.

The optical spectroscopy setup consists of a phase EOM to create sidebands for transient spectral holeburning spectroscopy and three AOM in series to produce laser pulses with a high extinction ratio. A polarization controller was used to select the polarization modes of the cavity. Two variable optical attenuators (VOA) were installed in the path, and 1 \% of the laser power was sent to a photodetector for power monitoring. The latter provides proportional voltage feedback to the VOA to stabilize the laser power. The stabilized laser power was routed to the fiber cavity mounted on the 10 mK stage of the dilution refrigerator. A 3-axis vector magnet was used to apply a tunable magnetic field for spectroscopy. The light collected from the fiber cavity was routed to a coupler where 1 \% power was sent to a high-gain photodetector (Femto OE-200-IN2) to monitor the cavity signal or a light trap (Thorlabs FTAPC1) during PLE measurements. The 99 \% of the power was directed towards a superconducting nanowore single photon detector (SNSPD) operating at 70 \% detection efficiency with a dark count of $\approx$ 2.1 Hz. The total photon collection efficiency of the setup was estimated as 

\begin{equation}{\mathrm{\eta_{total} = \eta_{cav} \eta_{fiber} \eta_{passive} \eta_{snspd} }}
\end{equation}

\noindent
where $\eta_{cav} = \kappa_e/\kappa$ is the collection efficiency from the cavity, which is equal to 4 \% for a cavity that is under coupled, $\eta_{fiber} = $ 51 \% is the total fiber transmission efficiency including mode-mismatch with the cavity, $\eta_{passive}$ = 72.4 \% accounts for the loss from couplers and other components and $\eta_{snspd}$ $\approx$ 70 \% is the efficiency of the SNSPD. Thus, the total photon counting efficiency of the setup $\eta_{total}$ is 1.9 \%. 
 
\begin{figure*}[t!]
    \centering
    \includegraphics[width=0.4\textwidth]{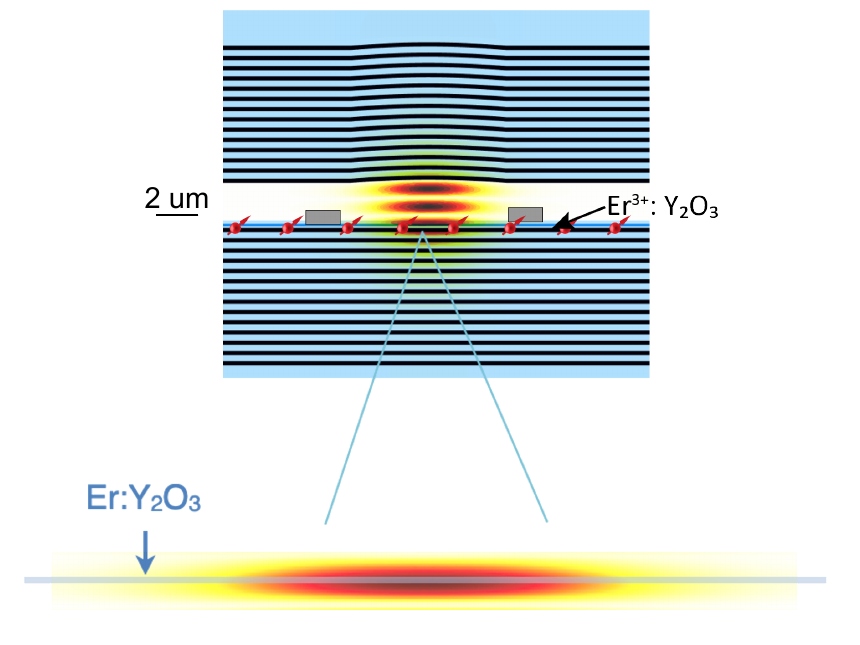}
    \caption[Fiber cavity mode profile]{ Optical fiber Fabry Perot cavity consists of a pair DBR mirror and is operated in $\mathrm{3\lambda/2}$ length. The optical mode simulated with COMSOL multiphysics \textregistered~\cite{comsol} shows maximum overlap with the $\mathrm{Y_2 O_3}$ thin film.}
    \label{fig:optical}
\end{figure*}

\subsection{Purcell enhancement and cavity mode volume}
Single $\mathrm{Er^{3+}}$ ion photoluminescence lifetime is Purcell enhanced from 8.5 ms to 0.14 ms (Fig.~4(a)) for C$_2$ site, which corresponds to a Purcell factor of 271.5 taking into account branching ratio ($\mathrm{\zeta}$ = 0.22) for the decay from $\mathrm{^{4} I _{13/2} }$ $\mathrm{Y_1}$ to $\mathrm{^{4} I _{15/2} }$ $\mathrm{Z_1}$ level \cite{Zhong2015ncomm},

\begin{equation}
    F_P= \frac{(\tau_0/\tau_c-1)}{\zeta} = \frac{(8.5/0.14-1)}{0.22}  = 271.4
\end{equation}

\noindent
The Purcell factor $\mathit{F_P}$ obtained here presents a lower limit on the maximum Purcell factor $\mathit{F_{P, Max}}$ due to misalignment between the dipole moment of $\mathrm{Er^{3+}}$ in the $\mathrm{C_2}$ sub-site and the cavity mode polarization. Nevertheless, this could be used to estimate the cavity mode volume ($\mathrm{V_m}$) following equation \ref{qbyv}. Substituting for the cavity quality factor Q, $\mathrm{F_P=271.4}$, wavelength $\mathrm{\lambda}$ and the local correction factor to the electric field \cite{PRLsinglesTian} $\mathrm{\chi_L= \frac{3n^2}{2n^2+1}}$, we get $\mathrm{V_m} = 8.9 \Big(\frac{\lambda}{n} \Big)^3 $.

\begin{equation}
    F_P = \frac{3}{4 \pi^2 \chi_L^2} \Big(\frac{\lambda}{n} \Big)^3 \frac{Q}{V_m}
    \label{qbyv}
\end{equation}

\noindent The mode volume can be used to obtain the single ion coupling strength g as \cite{PhysRevA.80.062307},

\begin{equation}
    g_0= \frac{\mu}{n} \sqrt{\frac{\omega_a}{2\hbar \epsilon_0 V_m}}
\end{equation}

\noindent
Subsituting $\mathrm{V_m} = 8.9 \Big(\frac{\lambda}{n} \Big)^3 $, $\omega_a/2\pi = 195,100$ GHz, and dipole moment for Er\textsuperscript{3+} Y\textsubscript{1}-Z\textsubscript{1} optical transition in C\textsubscript{2} site $\mu = \frac{3 \hbar e^2 n f}{2 m_e \omega \chi_L} =$ 9.7 $\times$ 10\textsuperscript{-33} Cm \cite{THIEL2011353}, we get a single ion coupling strength g$_0$=0.95 MHz. This is excellent agreement with the g$_0$=1.0 MHz obtained from $4g^2/\kappa=1/2\pi T_{1, {\rm {cav}}}$, where $T_{1, {\rm {cav}}}$ is the Purcell enhanced lifetime of 0.14 ms.

\subsection{Optical T\textsubscript{1} vs detuning in $\mathrm{C_{2}}$ site}

\noindent
The Purcell enhanced $\mathrm{T_1}$ depends on the detuning between the laser and the center of the cavity, following the expression,

\begin{figure*}[t!]
    \centering
    \includegraphics[width=0.4\textwidth]{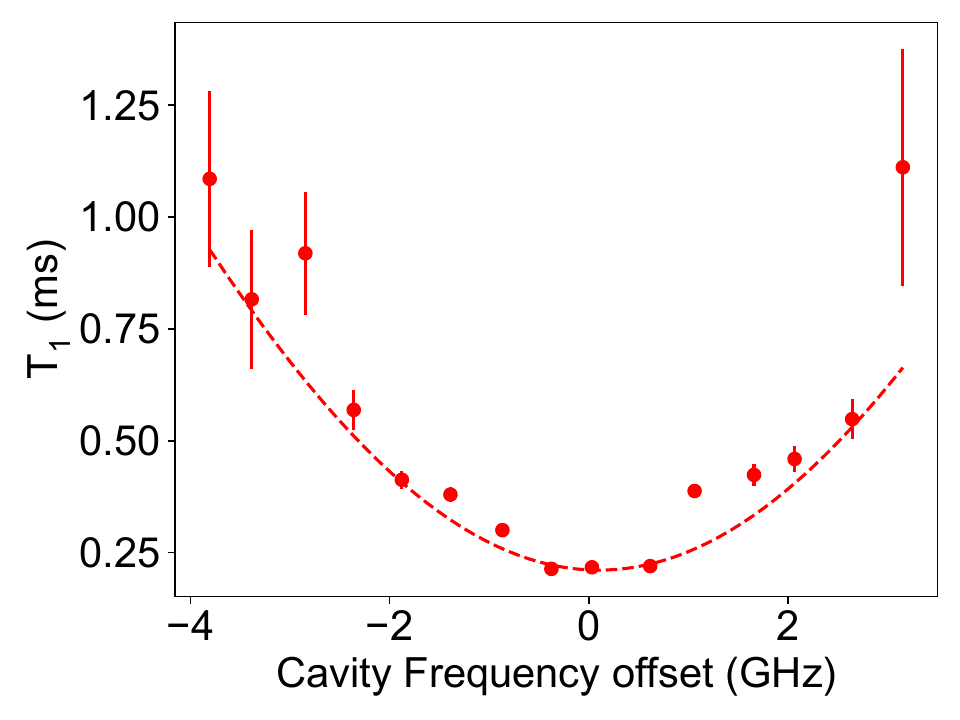}
    \caption[Optical T\textsubscript{1} vs laser detuning]{T\textsubscript{1} of $\mathrm{Er}^{3+}$ in $\mathrm{C_{2}}$ site as a function of detuning between the laser and cavity frequency. Varying the laser offset results allows tuning the Purcell enhancement factor and, therefore, the T\textsubscript{1}. Solid lines show a fit to Eq.~\ref{T1vsdetuning}.}
    \label{fig:c2t2detune}                                                                                                                                                                                                                                                                                                                                                                                                                                                                                                                                                                                                                                                                                                                                                                                                                                                                                                        
\end{figure*}

\begin{equation}
   \mathrm{ \frac{1}{T_{1,cav}} = \frac{1}{T_{1,0}} \Big( 1+  \zeta (F_P\frac{ (\gamma /2)^2}{((\gamma /2)^2 + \delta^2)} -1) \Big) }
   \label{T1vsdetuning}
\end{equation}

\noindent
where $\mathrm{T_{1,cav}}$ is the cavity-enhanced emitter decay, $\mathrm{T_{1,0}}$ is the intrinsic lifetime in absence of cavity enhancement, $\mathrm{\gamma}$ is the cavity linewidth, $\mathrm{\zeta}$ = 0.22 is the branching ratio, $\mathrm{F_P}$ is the measured Purcell enhancement and $\mathrm{\delta}$ is the laser detuning from the cavity center. Figure~\ref{fig:c2t2detune} shows the measured 1/$\mathrm{T_{1,cav}}$ as a function of the laser detuning from cavity center fitted to the model in equation~\ref{T1vsdetuning}.

\subsection{Optical homogeneous linewidth in $\mathrm{C_{2}}$ and C\textsubscript{3i} site}

\noindent
{\bf Optical homogeneous linewidth in C\textsubscript{2} site versus T\textsubscript{1}}
Application of magnetic field leads to a reduction of the optical linewidth along with a weak dependence on $\mathrm{T_1}$ as shown in Fig.~4(b) of the main text. The linewidth broadening trend can be modeled by a combination of broadening mechanisms,

 \begin{equation}
   \mathrm{  \Gamma_{eff} = \Gamma_0 + \Gamma_{SD,eff} ( 1- e^{-RT}) + \Gamma_{TLS} log(T_1/t)  }
   \label{gammeqn}
 \end{equation}
\noindent
where $\Gamma_0$ is the linewidth in short timescale (t$\mathrm{< 150 \, \mu s}$) including homogeneous linewidth, fast spectral diffusion, superhyperfine broadening, and drift of laser during the measurement timescale,  $\mathrm{\Gamma_{SD}}$ is the linewidth due to dipolar coupling to spin bath inducing spectral diffusion, R is the effective spin-bath flip rate and $\mathrm{\Gamma_{TLS}}$ is the dipolar coupling to electrical TLS. Further, $\mathrm{\Gamma_{SD, eff}}$ has an explicit dependence on the nature and splitting of the spin bath and can be described as,

\begin{equation}
   \mathrm{ \Gamma_{SD,eff} = \Gamma_{SD} \, sech^2 \Big(  \frac{g_{env} \mu_B B}{2 k_B T} \Big) }
\end{equation}
\noindent
where $\mathrm{\Gamma_{SD}}$ is the maximum dipolar coupling strength to the spin bath, $\mathrm{g_{sB}}$ is the effective g factor of the spin bath, B is the magnetic field, and T is the temperature. We assumed $\mathrm{g_{env}}$ =1.2, T= 100 mK (this combination is not unique and is chosen to obtain the best fit) to fit equation \ref{gammeqn} to obtain the parameters $\mathrm{\Gamma_{SD}}$ = 4.32 (1.1), $\mathrm{\Gamma_{0}}$ = 0.88 (0.11), $\mathrm{\Gamma_{TLS}}$ \textless  0.008 MHz and R = 0.83 (0.32) kHz. The fitted parameter values indicate spectral diffusion induced by paramagnetic spin bath as the primary source of spectral diffusion at B=0, which is frozen by application of B $\approx$ 250 mT. The contribution from TLS is \textless 0.01 MHz, indicating the absence of slow spectral diffusion from TLS. The short timescale linewidth $\mathrm{\Gamma_0}$ of 0.95 MHz indicates broadening mechanisms such as fast spectral diffusion by non-magnetic noise. A similar narrowing of homogeneous linewidth with magnetic was also observed for C\textsubscript{3i} site, as discussed below.\newline

\begin{figure*}[t!]
    \centering
    \includegraphics[width=0.4\textwidth]{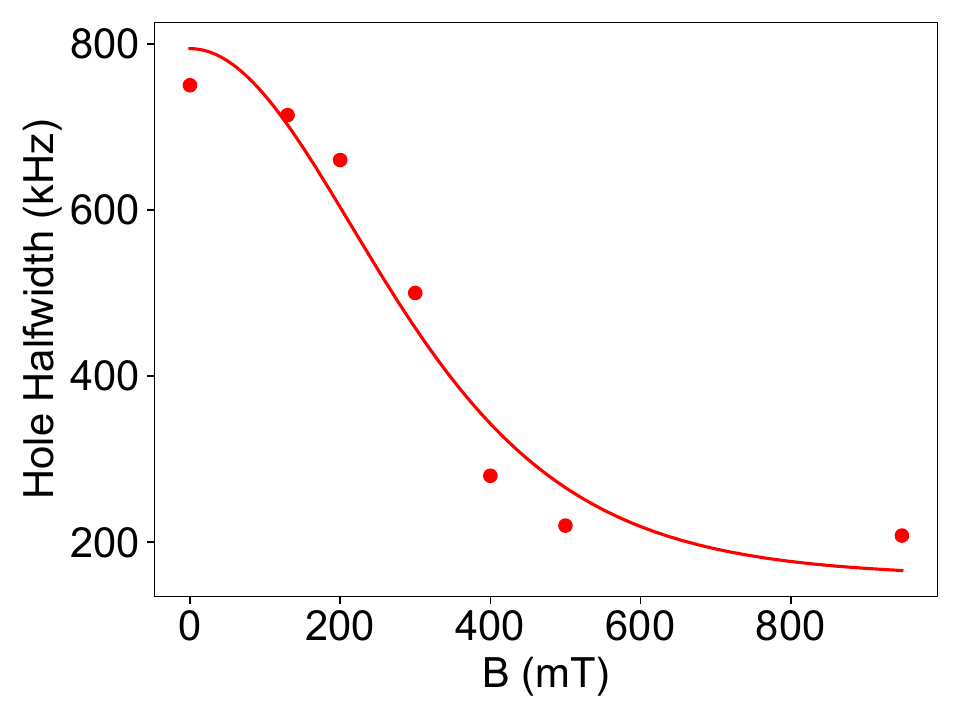}
    \caption[Homogeneous linewidth vs B field for C\textsubscript{3i} site]{Homogeneous linewidth for $\mathrm{Er}^{3+}$ in $\mathrm{C_{3i}}$ site as a function of magnetic field. Freezing of paramagnetic defects results in the narrowing of homogeneous linewidth with an applied magnetic field in close agreement with the model (solid lines).}
    \label{fig:c3iholewidth}
\end{figure*}

\noindent
{\bf Optical homogeneous linewidth in C\textsubscript{3i} site versus B field}
Homogeneous linewidth measured at 5 nW power using transient spectral holeburning (Fig.~\ref{fig:c3iholewidth}) shows a dependence on the magnetic field, which indicates the freezing of the spectral diffusion noise from a paramagnetic spin bath. The zero magnetic field linewidth of 800 kHz is in good agreement with the single ion linewidth measured at zero field and with the same power. The magnetic field dependence is fitted with a spectral diffusion-induced linewidth broadening model,

\begin{equation}
    \mathrm{\Gamma = \Gamma_0 + \Gamma_{SD} sech^2 \Big(  \frac{g_{env}\mu_B B}{2k_BT} \Big)}
\end{equation}

\noindent 
where $\mathrm{\Gamma_{SD}}$ is the dipolar linewidth due to coupling to a paramagnetic spin bath with an effective g-factor g$_{env}$ and $\Gamma_0$ is the magnetic field independent part of the linewidth, which includes broadening due to laser power, laser linewidth, instantaneous spectral diffusion. We obtain $\Gamma_0$= 158.6 (92.9) kHz, $\mathrm{\Gamma_{SD}}$= 635.5 (121.4) kHz, $\mathrm{g_{env}}$= 2.02 and T=0.22 K. The fitted parameters  $\mathrm{g_{env}}$ and T had a strong correlation (C=1), and they are not unique to obtain a good fit. In the future, this could be improved by performing separate measurements to estimate the sample temperature T. Nevertheless, the model suggests freezing of paramagnetic impurities as the mechanism of homogeneous linewidth narrowing at higher B field. 

\end{document}